\documentclass[sigconf]{acmart}
\AtBeginDocument{%
  \providecommand\BibTeX{{%
    \normalfont B\kern-0.5em{\scshape i\kern-0.25em b}\kern-0.8em\TeX}}}

\copyrightyear{2022}
\acmYear{2022}
\setcopyright{rightsretained}
\acmConference[ICSE '22]{44th International Conference on Software Engineering}{May 21--29, 2022}{Pittsburgh, PA, USA}
\acmBooktitle{44th International Conference on Software Engineering (ICSE '22), May 21--29, 2022, Pittsburgh, PA, USA}
\acmDOI{10.1145/3510003.3510172}
\acmISBN{978-1-4503-9221-1/22/05}

\usepackage{amsmath,amsfonts}
\usepackage{graphicx}
\usepackage{textcomp}
\usepackage[draft]{minted}
\usepackage{algorithm}
\usepackage[noend]{algpseudocode}
\usepackage{url}
\usepackage{booktabs}
\usepackage{subfigure}
\usepackage{multirow}
\usepackage{diagbox}
\usepackage{natbib}
\usepackage{lineno,hyperref}
\newcommand{\cf}{CodeFill}
\begin{document}

\title{CodeFill: Multi-token Code Completion
by Jointly Learning from Structure and Naming Sequences}

\author{Maliheh Izadi}
\email{M.Izadi@tudelft.nl}
\affiliation{
  \institution{Delft University of Technology}
  \city{Delft}
  \country{Netherlands}
}
\author{Roberta Gismondi}
\email{R.Gismondi@student.tudelft.nl}
\affiliation{
  \institution{Delft University of Technology}
  \city{Delft}
  \country{Netherlands}
}
\author{Georgios Gousios}
\email{G.Gousios@tudelft.nl}
\affiliation{
  \institution{Delft University of Technology}
  \city{Delft}
  \country{Netherlands}
}

\renewcommand{\shortauthors}{Izadi, et al.}

\begin{abstract}
Code completion is an essential feature of IDEs,
yet current autocompleters
are restricted to either grammar-based or
NLP-based single token completions.
Both approaches have significant drawbacks:
grammar-based autocompletion is restricted
in dynamically-typed language environments,
whereas NLP-based autocompleters
struggle to understand the semantics
of the programming language and the 
developer's code context.

In this work,
we present \textit{CodeFill},
a language model for autocompletion
that combines learned structure and naming information.
Using a parallel Transformer architecture and multi-task learning,
CodeFill consumes sequences of source
code token names and their equivalent AST token types.
Uniquely, CodeFill is trained both for single-token and multi-token
(statement) prediction, which enables it to learn long-range
dependencies among grammatical and naming elements.
We train CodeFill on two datasets,
consisting of $29$M and $425$M lines of code, respectively.
To make the evaluation more realistic, we develop a method to automatically
infer points in the source code at which completion matters.
We compare CodeFill
against four baselines and two state-of-the-art models,
\textit{GPT-C} and \textit{TravTrans+}.
\cf{} surpasses all baselines
in single token prediction
(MRR: $70.9\%$ vs. $66.2\%$ and $67.8\%$)
and outperforms the state of the art
for multi-token prediction
(ROUGE-L: $63.7\%$ vs. $52.4\%$ 
and $59.2\%$, for $n=4$ tokens).
We publicly release our source code and datasets.
\end{abstract}

\ccsdesc[500]{Software and its engineering~Software notations and tools}

\keywords{Automatic Code Completion,
Transformers,
Multi-Task Learning,
Types,
Dynamically-typed Languages}

\maketitle
\section{Introduction}\label{sec:intro}

Automatic code completion (also called autocompletion)
is the task of completing source code statements
by predicting what the developer would write
given the current context.
It helps developers finish their programming tasks faster by
decreasing the typing effort and saving keystrokes,
correcting typographical errors,
and enabling them to explore APIs
in a context-sensitive manner~\cite{amann2016study}.
Autocompletion has therefore emerged
as one of the most prominent features
in Integrated Development Environments (IDEs).

To support autocompletion,
current IDEs exploit the regular structure of programming languages.
For example,
an IDE knows that an opening parenthesis character (`\texttt{(}`)
at a function-call position
must be followed by enough arguments to match the function's arity.
It can therefore propose argument names for variables that are in scope.
The availability of types in the host programming language
helps increase the precision of suggestions;
continuing with the example above,
the IDE will only propose variable names
for variables whose types match the function argument.
Recent autocompletion systems also
take into account past completions~\cite{robbes2010improving}
and analyze large code bases~\cite{bruch2009learning}
to rank suggestions according to their past popularity.
Despite the best efforts of researchers and IDE developers,
developers find rule-based code completion mechanisms lacking.
Ranking suggestions based on alphabetical or usage frequency
(or even the suggestion list length~\cite{xianhao2018the})
neglects the current context,
leading to unrelated recommendations~\cite{allamanis2018survey}.
These problems are exacerbated in dynamically typed language settings,
as the IDE is lacking significant information to provide accurate suggestions.

To mitigate rule-based autocompletion issues,
researchers have proposed
statistical~\cite{nguyen2013statistical, hellendoorn2017deep} and
learning-based~\cite{liu2016neural,hellendoorn2017deep,li2017code,svyatkovskiy2019pythia,liu2020self,aye2020sequence}
autocompletion models.
Motivated by the \textit{naturalness} hypothesis~\cite{hindle2012naturalness},
learning-based models treat source code as natural language text,
hence code completion becomes an instance of the well-studied
text completion problem.
However, treating source code as text deprives learning-based models
of important code structure and semantic information~\cite{hellendoorn2019code}.
Moreover, the open-ended nature of code
leads to extremely large prediction spaces
due to developers constantly inventing identifier names~\cite{karampatsis2020big}.

In an illuminating study, Hellendoorn et al.~\cite{hellendoorn2019code}
identified a set of issues with current research in code completion.
Initially, the current approach of evaluating
accuracy as masked token prediction
does not reflect how autocompletion is used;
developers only trigger autocompletion after specific,
and certainly not arbitrary, points
in a program's syntax (e.g., after an opening parenthesis).
Thus, treating all tokens equally
masks the fact that some tokens (e.g., punctuation)
are much easier to predict than others (e.g., identifiers).
Moreover, most approaches (especially learning-based ones)
do not account for names coming from dependencies,
which deprives them of important context.

In this work, we propose \textit{CodeFill},
a novel learning-based
approach that aims to address the problems identified above.
\cf{} borrows from the bimodality hypothesis~\cite{santanu2018refinym}
to model source code inputs.
Specifically, \cf{} exploits that information is conveyed by
source code through two channels;
the \textit{natural language channel} (variable names, functions, etc.),
and the \textit{code structure channel} (inheritance, containment, etc.).
Inputs are fed into the model
simultaneously as both sequences of token values,
which enable it to learn relationships among token values,
and, uniquely, sequences of token types,
which enable it to learn associations between syntactic elements.
\cf{} is then asked to predict either the value or the type of the
next \emph{n} tokens.
To enable \cf{} to learn name dependencies across longer ranges,
we also train it with an additional task,
multi-token statement completion at the value level.
The input token names to \cf{} is encoded with Byte-Pair Encoding (BPE),
which enables \cf{} to both
compress the input name space and generate names that are not in the input vocabulary.
To present suggestions relevant to the developer's context,
\cf{} includes a post-processing step
that re-ranks the predictions based on
the context visible to the model at the completion point.
\cf{} is instantiated as a set of
three Transformers (GPT2-based)
trained with soft parameter sharing  Multi-Task Learning (MTL) setting.
Each transformer models one of the three tasks,
namely token value, token type, and multi-token prediction;
a joint loss function across all three tasks
updates the weights of all three model components.
During each epoch, the model is trained
on one task according to a configurable task-picking policy.
Our target language is Python, to both demonstrate the efficiency of the model when
type information is missing and also make our work comparable with the state of the art.

We pit \cf{} against four baseline models and two the-state-of-the-art models, namely
GPT-C~\cite{svyatkovskiy2020intellicode} and TravTrans+~\cite{kim2020code}.
We use two deduplicated datasets:
the ETH150K dataset (deduplicated: \textit{PY117K})
and a manually collected dataset
consisting of practically all
non-forked Python repositories on GitHub (\textit{PY1690K}).
We evaluate all models on two tasks:
\textit{Token-Level} and \textit{Statement-Level} Predictions (TLP and SLP).
For TLP, we evaluate for
i) next token prediction (TLP-A),
ii) next token type prediction (TLP-B),
iii) next token value prediction (TLP-C).
To ensure that the evaluation setting reflects real-world use of autocompletion,
we also evaluate completions after specific syntactic
elements, e.g., a dot \texttt{.} or an \texttt{AWAIT} keyword (TLP-D).
We devise an algorithm to identify those syntactic elements
(\emph{cardinal points}) automatically given a corpus.
We use top-1 Accuracy and the Mean Reciprocal Rank (MRR)
as evaluation metrics.
For the SLP task, we assess the models
on statement completion with $n$ tokens
and we compare them using METEOR and ROUGE-L measures.
To show that each component in the \cf{} model is necessary,
we perform an ablation study.

The results demonstrate that \cf{} outperforms all the competing approaches in all tasks.
Indicatively,
for each of the TPL-A, TPL-B, TPL-C, and TPL-D evaluation tasks,
\cf{} achieves a state of the art MRR of
$81.7\%$, $87.2\%$, $69.5\%$, $70.2\%$
while TravTrans+, a current state of the art, scores
$79.4\%$, $83.6\%$, $63.8\%$, and $66.2\%$,
respectively.
In the SLP evaluation task,
for completing statements with four tokens
(the average completion length in our datasets)
\cf{} obtains $70.2\%$ and $63.8\%$
for the METEOR and ROUGE-L metrics respectively,
and thus significantly surpasses TravTrans+
($64.5\%$ and $52.4\%$).

The main contributions of this work are:
\begin{itemize}
    \item \cf{}, a model that unifies learning of structural and name-based information for the autocompletion task.
    \item An implementation of \cf{}, including training procedures,
    for the Python programming language.
    We make our code and datasets available.~\footnote{\url{https://github.com/saltudelft/codefill}}
    \item An extensive evaluation of \cf{} 
    against four baseline models and two state-of-the-art approaches, demonstrating its superior performance.
\end{itemize}

\section{Background and Related Work}\label{sec:bg}
In this section,
we briefly review the background work relating to our approach.
Then, we present the main approaches to autocompletion,
including the baselines we used for comparison.

\subsection{Language Models and Transformers}
\label{sec:bg_transformers}
Statistical Language Modeling (LM)
is the task of developing a probabilistic model for predicting
the next tokens in a sequence given its preceding tokens,
i.e., the context~\cite{goldberg2017neural}.
This context for simpler LMs is a short sequence of words,
while it can be sentences or paragraphs for larger models~\cite{schutze2008introduction}.
LMs are either used without modification,
e.g., in a text generation task,
or used inside a down-stream task
which requires language understanding.
Programming languages also
contain predictable statistical properties
which can be learned using LMs~\cite{hindle2012naturalness}.

Recently, Neural LMs have gained popularity
due to their superior performance and generalization capabilities~\cite{mikolov2010recurrent,goldberg2017neural}.
Neural LMs address the n-gram data sparsity problem
through parameterization of words as vectors~\cite{kim2016character}.
A real-valued vector (word embedding) is used to
represent each word in a vector space.
This representation of words
is learned based on their usage.
This allows words with a similar meaning
to have a similar representation.
Note that traditional statistical LMs
were not able to achieve this level 
of generalization~\cite{schwenk2002connectionist}.
Moreover, the distributed representation approach
makes it easier for the embedding representation to scale
with the vocabulary size.
This is specifically useful with source code,
where the vocabulary size can be unlimited
due to the use of arbitrary identifiers.
Initially, feed-forward neural network models,
then Recurrent Neural Networks (RNNs)
and next, networks with long-term memory,
such as Long Short Term Memory (LSTM) networks were used.

Most recently, there have been significant improvements
with the introduction of self-attention architectures in the Transformer
which is a sequence-to-sequence architecture
for transforming a given sequence of elements to another form~\cite{vaswani2017attention}.
Attention enable Transformers
to focus on selective parts of an input,
thus generating more relevant outputs~\cite{luong2015effective}.
Transformers outperform previous deep models
such as RNNs and LSTMs on multiple NLP tasks~\cite{vaswani2017attention}.
A Transformer consists of two main components,
an encoder, and a decoder.
GPT-2 introduced by OpenAI~\footnote{\url{https://openai.com/}},
is a large generative Transformer-based LM
trained on a dataset of $8$M web pages~\cite{radford2019language}.
GPT-2 has been successfully exploited
for various NLP and source code analysis tasks~\cite{budzianowski2019hello,lee2020patent,ham2020end,svyatkovskiy2020intellicode}.

\subsection{Multi-Task Learning}\label{sec:bg_mtl}
Multi-Task Learning (MTL) is a model training technique that combines
multiple tasks and a joint loss function,
with the goal of maximizing performance on one or all of the tasks.
MTL enables knowledge transfer across related tasks
and improves generalization by leveraging the domain-specific
information contained in the training signals of related
tasks~\cite{caruana1997multitask}.
An MTL model captures the common features among all the tasks
through sharing hidden layers among them.
MTL has been applied successfully in both NLP~\cite{devlin2018bert}
and source code analysis~\cite{liu2020self,liu2020multi}.
There are two approaches to jointly train models using MTL,
\textit{hard-parameter} and \textit{soft-parameter} sharing.
In the former, the hidden layers
are shared between all tasks
while keeping several task-specific output layers.
For the latter,
each task has its own model with its own parameters.
However, the distance between them is regularized
to encourage the parameters to be similar.
In the soft-parameter sharing case, training can happen
either \textit{sequentially} (one task per training round)
or \textit{alternatively} (one task per epoch).

\subsection{Related Work}\label{sec:related}
Autocompletion is an active research area
for both practitioners and researchers.
Below, we review the latest approaches to autocompletion.

\subsubsection{Conventional Autocompletion}
Traditionally, autocompleters used
heuristic rules
static type information~\cite{hou2010towards},
similar code examples~\cite{bruch2009learning},
and program history data~\cite{robbes2008program}
for suggesting completions.
For instance, IDEs conventionally return
a list of type-checked names
either based on the order of alphabet or usage frequency.

\subsubsection{Statistical LMs and Grammar-based Models}
Several studies use statistical LMs for modeling source code~\cite{hindle2012naturalness,nguyen2013statistical,tu2014localness,hellendoorn2017deep}.
Tu et al.~\cite{tu2014localness} built upon an n-gram model
using a cache mechanism to capture locality in source code.
Hellendoorn and Devanbu~\cite{hellendoorn2017deep} improved the n-gram model
by exploiting various techniques including
nested scopes, locality, and unlimited vocabulary.
Raychev et al.~\cite{raychev2016probabilistic}
proposed a probabilistic model
based on decision trees and domain-specific grammars.
Researchers also studied the use of syntactic structure
through exploiting probabilistic graphical models.
Allamanis et al.~\cite{allamanis2014mining}
employ probabilistic context-free grammars,
while Raychev et al.~\cite{raychev2016probabilistic,raychev2016learning,bielik2016phog}
use probabilistic higher order grammars to this end.

\subsubsection{Deep Learning for Autocompletion}
Recently, deep neural networks
such as RNNs, LSTMs and Transformers
are being effectively used
for modeling source code~\cite{li2017code,hellendoorn2017deep,aye2020sequence,karampatsis2020big,kim2020code}.
In 2018, Li et al.~\cite{li2017code} proposed
a pointer mixture model to mitigate the Out-Of-Vocabulary (OOV) problem.
They trained two LSTM models on types and tokens.
Karampatsis et al.~\cite{karampatsis2020big}
presented a large-scale open-vocabulary neural LM.
They incorporated BPE, beam search, and cache mechanism
to address the OOV problem.
Most recently, Kim et al.~\cite{kim2020code},
incorporated the syntactic structure of trees
into their Transformer-based model to better learn
from source code.

\subsubsection{Multi-token Autocompletion}
Although most research on code completion
is focused on single-token prediction,
several studies aimed to complete entire statements or blocks of code~\cite{svyatkovskiy2020intellicode,nguyen2019combining,yang2017language,wen2021siri}.
Yang et al.~\cite{yang2017language}
proposed \textit{PCC} and
introduced an intermediate representation for source code,
to put tokens into groups using lexeme and variable relative order.
Nguyen et al.~\cite{nguyen2019combining}
proposed \textit{AUTOSC}
to combine program analysis and software naturalness
and fill in a partially completed statement
with frequent and valid recommendations.
Svyatkovskiy et al.~\cite{svyatkovskiy2020intellicode}
recently proposed a GPT-2 based multi-lingual model,
\textit{GPT-C}, for completing lines.
Wen et al.~\mbox{\cite{wen2021siri}}
introduced \textit{FeaRS} which recommends the next method
given the current code in an IDE
using implementation patterns
learned through mining open source projects.

\subsubsection{MTL for Autocompletion}
MTL has been used in various NLP-related tasks~\cite{zhang2021survey,ruder2017overview,sener2018multi}.
Recently, it has also been employed
for programming language processing tasks.
Liu et al.~\cite{liu2020multi,liu2020self}
proposed two approaches based on MTL for autocompletion.
In the first study, the authors used a Transformer-XL and an RNN
for predicting next token type and value~\cite{liu2020self}.
They develop a partial AST encoder and a path2root encoder 
and use them in their MTL framework. 
In their second study,
Liu et al.~\cite{liu2020multi} 
pre-train their model with hybrid objective functions
for code understanding and code generation tasks. 
Next, they fine-tune it on code completion.
The pre-training tasks are masked bidirectional LM, 
next code segment prediction, and unidirectional LM.
The fine-tuning tasks are unidirectional masked LM, 
and unidirectional LM.

\subsubsection{Practical Aspects of Autocompletion}
Hellendoorn et al.~\cite{hellendoorn2019code} claim
the accuracy of autocompleters evaluated on synthetic data
can drop on real-world data.
Aye et al.~\cite{aye2021learning},
trained models on real-world code completion examples
of an internal dataset (Facebook).
They showed that models trained on data distributions that
are closer to those of where the model will be deployed
can outperform models trained on
committed source code in public repositories.
Svyatkovskiy et al.~\cite{svyatkovskiy2019pythia}
integrated Pythia, an LSTM model, to \textit{IntelliCode},
an extension to Microsoft VS Code IDE.
In a follow-up study~\cite{svyatkovskiy2020intellicode},
they introduced \textit{IntelliCode Compose}
as a general-purpose multilingual autocompletion using Transformers.
The improved model predicts sequences of code tokens,
generating up to entire statements.
IntelliCode Compose is integrated into the Microsoft VS Code IDE.
Finally, Svyatkovskoy et al.~\cite{svyatkovskiy2021fast} implemented
and evaluated several neural code completion models, which
offer varying trade-offs in terms of memory, speed, and accuracy.
Commercial autocompletion tools,
such as \textit{TabNine} and \textit{GitHub Copilot} also exist,
but very little technical information has been shared about them.

\subsection{Baselines}
We include six
recent models as baselines
to provide a comprehensive evaluation.
For all baselines,
we use the replication packages provided by the authors
and set the parameters as defined in each respective study.
For the statement level prediction task,
we modified the output layer of the baselines
to predict up until the end of a statement.

\textbf{N-gram + LSTM (FSE, 2017)}:
Hellendoorn et al.~\cite{hellendoorn2017deep}
claim that a well-engineered and simple approach
(n-gram based language models)
can provide better performance than more complex models (deep neural networks).
The authors show that the combination of an n-gram and LSTM-based model outperforms the rest of their models.

\textbf{Pointer Mixture (IJCAI, 2018)}:
Li et al.~\cite{li2017code},
propose a pointer mixture model
to address the OOV problem.
They also try to incorporate structural information in their models
by training two models (token types and values) separately.

\textbf{T-XL + Bi-LSTM (ICPC, 2020)}:
Liu et al.~\cite{liu2020self,liu2020multi},
propose two models based on the MTL technique.
The first study uses Transformer-XL and a Bi-LSTM
to train two models for tokens and AST paths
for dynamically-typed languages such as Python.
The second study by the same group
presents a pre-trained language model
which is fine-tuned for code completion.
The authors use static analysis and type annotations
for their type prediction task, for Java.
We compare against the first model only,
as it most closely matches our setup.

\textbf{OpenVocab (ICSE, 2020)}:
To address the OOV problem,
Karampatsis et al.~\cite{karampatsis2020big}
present a BPE-based language model.
We include it here for completeness,
even though their model is not tuned for autocompletion.

\textbf{IntelliCode Compose (FSE, 2020)}:
Svyatkovskiy et al.~\cite{svyatkovskiy2020intellicode}
propose a general-purpose, multi-lingual autocompletion
supporting multi-token statement completion.
They train a GPT-2 model on $1.2B$ LOC
written in Python, C\#, TypeScript, and JavaScript.
This tool is deployed as a cloud-based web service
and uses client-side caching and parallel implementation
to speed up the predictions.
As the source code
is not publicly available,
we trained a GPT-2 model for source code
and did our best to adhere to the settings reported in the study.
As the focus of our study is mono-lingual,
we only train this model on Python code.

\textbf{TravTrans+ (ICSE, 2021)}:
Kim et al.~\cite{kim2020code}
propose a transformer-based approach
which exploits AST paths.
We use their best model, TravTrans+,
as the state of the art in our evaluation.

\section{Approach}\label{sec:approach}
\begin{figure}[tb!]
  \centering
  \includegraphics[width=\linewidth]{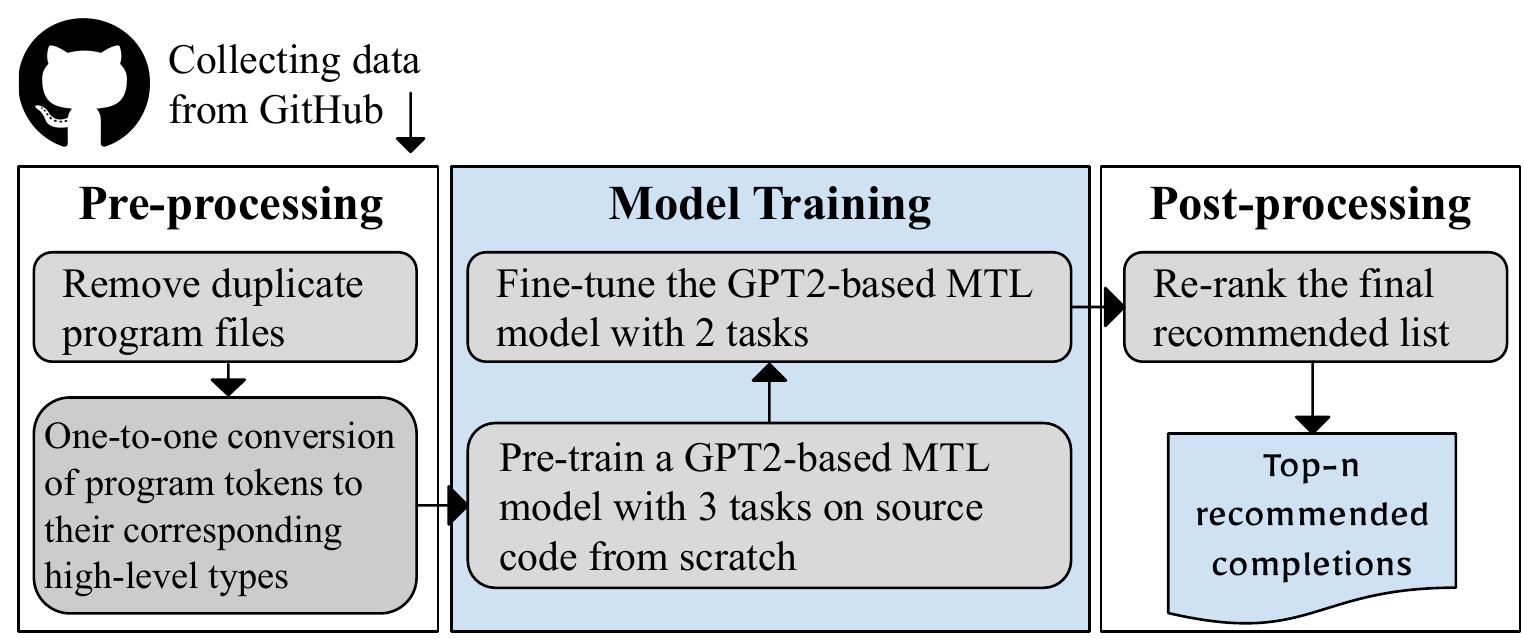}
  \caption{\cf{} Workflow}
  \label{fig:workflow}
\end{figure}

The \cf{} pipeline comprises two main phases;
\textit{pre-processing},
\textit{model training}.
Figure \ref{fig:workflow} presents the overall workflow.
Initially, \cf{} pre-processes, tokenizes
and converts the input source code 
to equivalent syntax token sequences.
Training consists of two main phases
\textit{pre-training} with 3 tasks (token sequence type and name completion, statement completion)
and \textit{fine-tuning} on 2 tasks (name and statement completion).
For both stages, \cf{} uses
soft-parameter sharing MTL
to learn from different representations of source code.
At evaluation time, Codefill also
re-orders recommendations
based on their type and the visible context.

In the following section, we present how the proposed approach works in detail.

\subsection{Pre-processing}\label{sec:approach_link}
During pre-processing, \cf{}
converts the input program files
to an equivalent format where keywords and identifiers
are swapped with their AST equivalents.
The algorithm starts by removing comment sections, blank spaces, and blank lines.
It then extracts the list of
\textit{modules}, \textit{libraries}, and their \textit{aliases}
using the Python \textit{AST} library.
Those are stored in a dictionary
and, using it, \cf{} replaces all their occurrences in code
with their respective types
(i.e., \texttt{MODULE}, \texttt{LIBRARY}, and \texttt{ALIAS}).

\cf{} also pre-processes and tokenizes the input source code.
For each line, it reads the tokenized information
and stores four types of information about each token
namely
(1) its value,
(2) its type,
(3) its line number,
and (4) its position in the line.
For instance, for the statement \texttt{return node}
in Figure~\ref{fig:sample_code},
it stores two tokens
as shown in the table following the code example.
Moreover, variable visibility information
(e.g., global vs. local variables),
is maintained, to differentiate
between different name usages in the same context.
\begin{figure}[tb]
\centering
\begin{minted}[
linenos,frame=single,
breaklines=true,
framerule=0.5pt,
breaksymbolleft=,
numbersep=1mm,
fontsize=\footnotesize]{python}
def transform(node, ctx):
    node = qual_names.resolve(node)
    node = CallTreeTransformer(ctx).visit(node)
    return node
\end{minted}
\begin{tabular}{cccc}
    Type & Value & \#Line & Position \\\midrule
    RETURN &	return	& 4	& 1 \\
    NAME   & 	node	& 4	& 2  \\
    \bottomrule
\end{tabular}
\caption{Sample code snippet and the extracted information}
\label{fig:sample_code}
\end{figure}
%

To address the OOV problem,
\cf{} uses a BPE-encoded name representation.
Exploiting word segmentation,
BPE iteratively merges the most frequently occurring character sequences.
Prior to applying BPE encoding,
and similarly to other studies~\cite{svyatkovskiy2020intellicode,izadi2021topic,izadi2022predicting},
\cf{} normalizes the input strings
by replacing
\textit{string}, and \textit{numeric} literals
with respective special tokens,
i.e., \texttt{STRING} and \texttt{NUMBER}.

A unique characteristic of the Python language
is that indentation defines code blocks;
it is therefore important for source code models
to learn to encode indentation as part of their learned representation.
To do so, \cf{} stores the positioning of indentation markers.
For the first line with an indentation,
it adds a special token $\langle INDENT \rangle$
at the beginning of the given line.
It passes through the following lines with the same indentation,
to reach the next indentation or a dedentation position,
at which point it adds a respective $\langle INDENT\rangle$
or $\langle DEDENT\rangle$  token.

The pre-processing step results in two files for each input source code file;
(1) one containing sequences of token names
minus the comments and extra blank lines, and
(2) one containing sequences of token types.
Both are fed into \cf{}
as two different but corresponding representations of source code.
Figure~\ref{fig:cherry} shows a sample function
and its corresponding type information with the correct indention.

\begin{figure}[tb!]
\centering
\begin{minted}[
linenos,frame=single,
breaklines=true,
framerule=0.5pt,
breaksymbolleft=,
numbersep=1mm,
fontsize=\footnotesize]{python}
# Raises an error when the required variable is missing
def required_env(var):

    value = os.environ.get(var)
    if value is None:
        raise RuntimeError("Var is required to start the service.")
    return value
\end{minted}
\begin{minted}[
linenos,frame=single,
breaklines=true,
framerule=0.5pt,
breaksymbolleft=,
numbersep=1mm,
fontsize=\footnotesize]{python}
def required_env(var):
    value = os.environ.get(var)
    if value is None:
        raise RuntimeError("STRING")
    return value
\end{minted}
\begin{minted}[
linenos,frame=single,
breaklines=true,
framerule=0.5pt,
breaksymbolleft=,
numbersep=1mm,
fontsize=\footnotesize]{python}
DEF FUNCTION_NAME(LOCAL_VARIABLE): EOS
INDENT LOCAL_VARIABLE = LIB.MODULE.FUNCTION_NAME(LOCAL_VARIABLE) EOS
IF LOCAL_VARIABLE IS NONE: EOS
INDENT RAISE ERRORTOKEN("STRING") EOS
DEDENT RETURN LOCAL_VARIABLE EOS
\end{minted}
\caption{An example code snippet and its converted version}\label{fig:cherry}
\end{figure}

\begin{figure*}[tb!]
  \centering
  \includegraphics[width=0.9\linewidth]{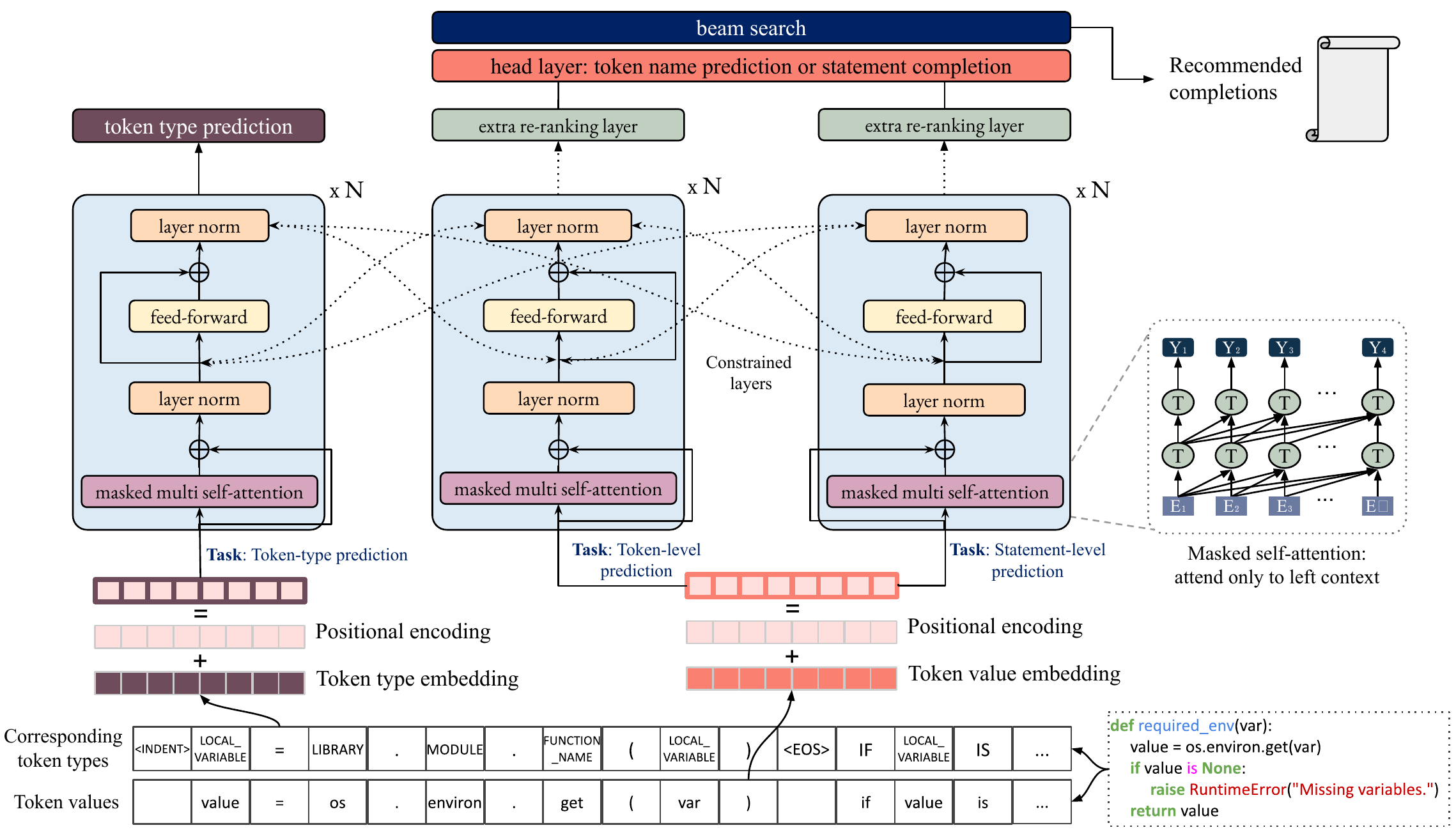}
  \caption{Model training}
  \label{fig:approach}
\end{figure*}

\subsection{Model Training}\label{sec:approach_modeltrainingl}
In this phase,
\cf{} learns from two granularity levels;
token- and statement-level completions
with three simultaneous tasks,
namely
(1) next Token \textit{Value} Prediction (TVP),
(2) next Token \textit{Type} Prediction (TTP),
and (3) \textit{Statement} Completion (SC).
Model training follows a two-stage process;
First, a generic language modeling objective
is used on the unlabeled data
to learn the initial parameters.
Then, these parameters are adapted
to the target tasks using the corresponding objective.
Thus, while pre-training,
\cf{} learns from all three tasks
while fine-tuning is restricted to
the TVP and SC tasks.
The reason for excluding the TTP task is that
the number of types for all the program files is limited.
Hence, the model quickly learns how to properly predict these type sequences
(i.e., learns an effective representation of the Python grammar),
eliminating the need for further fine-tuning.

The main neural network architecture
for all tasks is based on the GPT-2 Transformer with $L$ layers.
\cf{} uses three distinct GPT-2 transformers,
each with its own input and training objective.
The models are initialized with random weights.
Transformer blocks include
self-attention layer,
feed-forward neural nets,
and a normalization layer.
Self-attention blocks identify which tokens to focus on.
Feed-forward neural nets consist of an input layer to accept information,
hidden layers to capture the hidden correlations between each data point,
and finally, an output layer to transmit information.
The parameters are transferred to the next decoder
in the stack after being regularised (with $l2$ norm)
to be similar to the respective decoder's parameters.
\cf{} uses softmax activation function in the output layer
to generate probability distributions
over the vocabulary.

To train the model to predict a sequence of tokens,
$\{v_t\} \subset D, t \in [1, \dots, N]$,
with  $D$ as the vocabulary,
and $C$ as the existing code context,
\cf{} estimates the following conditional probability distribution, $P$:
\begin{equation}
    P(v_0,\dots,v_N|c_0,\dots,c_T) = \prod_{i=1}^{N}{P(v_i | c_0, \dots,c_T,v_0,\dots,v_{i-1})}.
    \label{eq:prob}
\end{equation}

We use a standard language modeling objective, 
predicting the next token given a context,
and maximize the following likelihood
based on our unsupervised corpus of tokens.
In Equation~\ref{eq:log_statement},
$m$ is the length of the predicted sequence of code token values
and $\theta$ is the set of parameters
that is learned through
stochastic gradient descent optimization to model $P$~\cite{robbins1951stochastic}.
\begin{equation}
    L(V) = \sum_{i}{\log{P(v_i | c_0,...,c_T,v_{i-m},...,v_{i-1};\theta)}}.
    \label{eq:log_statement}
\end{equation}

In each layer, multi-attention heads are used
to aggregate the output of the previous layer
for each transformer block.
Multi-headed self-attention is applied over the
input context tokens followed by position-wise feed-forward layers
to produce the output distribution.
\begin{equation}
h_0 = C W_e + W_p,
    \label{eq:attention1}
\end{equation}
\begin{equation}
h_l = transformer\_block(h_{l-1}), l \in [1, \dots, L],
    \label{eq:attention2}
\end{equation}
\begin{equation}
P(v_t) = softmax(h_L W_e^T), t \in [ 0,\dots,N]
    \label{eq:attention3}
\end{equation}
where $C$ is the context vector of tokens,
$L$ is the number of layers,
$W_e$ is the token embedding matrix,
and $W_p$ is the position embedding matrix.

For training with MTL,
\cf{} uses the alternative training strategy,
which aims to prevent catastrophic forgetting
(as opposed to the sequential strategy).
With a probability of
20\%, 40\%, and 40\% for each of the
TTP, TVP,
and SC tasks, respectively,
\cf{} picks a random task for each epoch.
TTP requires fewer epochs
as its vocabulary is fairly limited.
Further on, for TVP and SC tasks,
\cf{} uses beam search
to identify the most likely (sub-)token sequences.

Loss is shared among all tasks.
During pre-training, the parameters are tuned
to minimize the absolute  minimum of the cross entropy losses
among the three pre-training
tasks, namely, TVP, TTP, and SC (Equation~\ref{eq:shared_loss}).
When fine-tuning, only TVP and SC losses are used.
\begin{equation}
Loss_{final} = \mid\min({Loss_{TVP}, Loss_{TTP}, Loss_{SC}})\mid
\label{eq:shared_loss}
\end{equation}

\subsubsection{Token Value Prediction Task (TVP)}
\cf{} uses different representations of programs for each task
within the soft-parameter sharing MTL framework.
\cf{} treats the TVP task as masked unidirectional prediction;
left-side context is used to predict the next token.
The inputs to the task are sequences of token values,
represented as real-valued vectors of $[v_1, v_2, \dots, v_n]$.

\subsubsection{Token Type Prediction Task (TTP)}
Similarly to TVP, TTP is also treated as left-to-right
masked unidirectional prediction.
The input are corresponding token type representations
as real-valued vector of $ [t_1, t_2, \dots, t_n]$
As both the TTP and TVP models are trained jointly,
\cf{} is capable of exploiting token types 
when the ultimate goal is to predicting token values.

\subsubsection{Statement Completion Task (SC)}
As useful as next-token prediction may be,
developers can also benefit from
getting longer suggestions to complete code statements~\cite{svyatkovskiy2020intellicode,aye2020sequence,nguyen2019combining}.
Correspondingly, \cf{} can also benefit
from training to predict longer sequences,
as training will enable it to better prioritize context use.
Thus, we add a third task
to train \cf{} to provide completion suggestions
up and until the end of a statement.
To predict a whole statement
given the existing code context $C$,
and the vocabulary $D$,
\cf{} attempts to generate token values
$\{v_t\} \subset D$,
conditioned on the sequence of preceding token values
$\{c_t\} \subset D$,
For this task, the pre-processing steps introduce a special token ($\langle EOS \rangle$)
to demarcate the end of a statement.
\cf{} is trained to
keep predicting sequences of token names
until it produces an $\langle EOS \rangle$ token.

\subsubsection{Beam search}
\cf{} uses greedy (beam) search to identify the most probable sequences
given a sequence of probabilistic predictions.
Specifically, $ |B| $ (width of the beam) top probabilities,
are recorded partially for every step.
This heuristic algorithm does not necessarily optimize results;
however, its computational complexity equals to $ O(|B| \times |V|) $
which is much faster than computing all cases.
As $ |B| $ increases, the quality of generated summaries improves,
however, the learning time increases as well.
We experimented with several beam values (3, 5, and 10),
and settled to $5$, as it provided a good balance of accuracy and speed.

\subsection{Post-processing}\label{sec:appraoch_post}
\paragraph{Re-ranking Recommendations}
For a recommendation system to be useful,
predictions should be ranked similarly to user expectations.
To optimize ranking,
\cf{} includes a post-processing layer to re-rank the leaf nodes in the final
recommendation list based on the visible scope
(i.e., the current file).
This is based on the observation that most completions should be local to the edited file,
as naming visibility rules should force names to cluster.

To re-rank the suggestions,
\cf{} hierarchically divides the visible scope
to file, class, and closest function.
The intuition here is,
when the model is predicting the next token
and its type is expected to be a variable name,
candidates in the closest scope
have a higher probability of being correct.
However, when the next token
is predicted to be a function name,
candidates from the same class
(functions defined in the same class)
should be probably at the top of the list.
The re-ranking process consists of
multiplying the prediction probabilities of the top-10 predictions
with a corresponding weight coefficient.
The weights are selected based on
the type of the predicted token
and the scope of the declaration of the identifier.
Each prediction consists of a
\texttt{<token, type, probability>} triplet
with respect to the prediction point
that it is made available,
We generate the list of all visible names
and their hierarchical scope (function, class, file).
Each prediction is then cross-checked with this list,
in the case where the predicted identifier
is indeed already declared in the file (and thus in the list),
its prediction probability is multiplied by a weight
depending on the type of the predicted token
and the scope associated with the item in the list.
As the weights impact the quality of predictions,
we first defined a range/ratio
for different categories based on our programming intuition.
Then, we experimented with this range
and selected the best performing weights.
Table~\ref{tab:reranking-weights} presents
the weights used in this process.

Although the current weights
improve the predictions,
this only sets the minimum bar.
Future work can exploit automatic learning of these weights.

\begin{table}[tb!]
\caption{Weights in the post-processing layer for re-ranking}
\begin{small}
\centering
\begin{tabular}{cccc}
\toprule
    Leaf node type         & Function & Class & File \\
    \midrule
    Attribute Access  & 1.625 & 1.250 & 1.125 \\
    Variable names    & 1.625 & 1.125 & 1.500 \\
    Function names    & 1.125 & 1.625 & 1.500 \\
    \bottomrule
\label{tab:reranking-weights}
\end{tabular}
\end{small}
\end{table}

\begin{algorithm}[tb!]
\caption{Re-ranking final recommendations}
\label{alg:reranking}
\begin{algorithmic}[1]
\State \textbf{input} Predictions, WeightsList
\State \textbf{output} Predictions \Comment{List of updated predictions}
\State Names $\gets$ getSignificantNames() \Comment{Get the list of important names in left context from the  file}
\State \textbf{for} pred \textbf{in} Predictions \textbf{do}
\State \hskip1em \textbf{while} \textit{true} \textbf{do}
\State \hskip2em Names $\gets$ getSignificantName.pop()
\State \hskip2em \textbf{if} significantName.token = prediction.token \textbf{then}
\State \hskip3em typeCategory $\gets$ getTypeCategory()
\State \hskip3em weight $\gets$ weights[typeCategory][scope]
\State \hskip3em pred.probability $\gets$ pred.probability $\times$ weight
\State \hskip3em \textbf{break}
\State \hskip2em \textbf{end if}
\State \hskip1em \textbf{end while}
\State \textbf{end for}
\end{algorithmic}
\end{algorithm}

\section{Experimental Setup}\label{sec:experiments}
To train and evaluate \cf{},
we use two Python datasets.
We evaluate the models based on different evaluation scenarios,
to achieve a more realistic and comprehensive outlook
on the performance of code completion models
to benefit developers in real-world cases.

\subsection{Evaluation Tasks}\label{sec:eval_tasks}
We evaluate \cf{} on two tasks, namely
\textit{Token-Level} and \textit{Statement-Level} Predictions (TLP and SLP).

\subsubsection{Token-Level Prediction}
We use TLP to assess the ability of the model
to predict a single next token. We split this part of the evaluation
into four subtasks presented below.

\paragraph{Any token prediction} Our first sub-task is to evaluate the predictions
of any token irrespective of its type (TLP-A).
This is the baseline evaluation task employed in the literature,
but as research has shown~\cite{hellendoorn2019code},
it is not representative of
real-world autocompletion use. For this reason, we resort to more detailed
evaluations, as presented below.

\paragraph{Token Type Prediction} To assess the model's ability to learn grammatical sequences,
we evaluate how well a model can predict a correct AST token given
a context (TLP-B).
We group together AST tokens in the following categories:
\textit{Identifiers},
\textit{Keywords},
\textit{Operators},
\textit{Punctuation},
and finally numerals and string \textit{Literals}.

\paragraph{Leaf Node Prediction} Inspired by the evaluation setup of the state-of-the-art study by Kim et al.~\cite{kim2020code},
we investigate the ability of models when predicting AST leaf nodes (TLP-C),
including
\textit{Attribute access},
\textit{Names},
\textit{Function parameters},
and \textit{Constants}.

\paragraph{Cardinal Point Prediction}
The three tasks presented up to now give a comprehensive view
of the prediction ability of a model.
However, in practical settings, autocompletion is only triggered
at specific points
(e.g., after a dot,
or after specific keywords such as \texttt{for})
while the developer is editing source code.
To ensure that predictions translate to practical benefits
for the developers,
we evaluate completions on cardinal points (TLP-D).
To obtain a list of keywords
after which autocompletion is likely to be triggered,
we first select the list of
punctuation and keywords tokens that can be completed.
We then compute the frequency of
all bi-grams with any of these tokens as their first token
in our dataset.
Then, we remove three sets of bi-grams;
(1) those that are mostly written together 
with occurrence frequency above $95\%$ (e.g., \texttt{async def}),
(2) those that are normally not predictable (e.g., \texttt{class NAME} or \texttt{def FUNCTION-NAME}),
and finally (3) those that are usually not practical completions (e.g., \texttt{TRUE :}).
The resulting list of tokens
after which it is most beneficial for autocompletion to be triggered
is as follows.

\vspace{1em}
\noindent\fbox{
    \parbox{0.47\textwidth}{
        DOT, AWAIT, ASSERT, RAISE, DEL, LAMBDA, YIELD, RETURN,
        EXCEPT, WHILE, FOR, IF, ELIF, ELSE, GLOBAL, IN,  AND,
        NOT,   OR,  IS,  BINOP,  WITH, ;, ,, [, (, \{,  \textasciitilde
    }
}

\paragraph{Evaluation Metrics}\label{sec:TLP_metrics}
As the model only predicts a single token in the TLP task,
we include two evaluation metrics, namely
the \textit{Accuracy} of the top prediction
and the Mean Reciprocal Rank (\textit{MRR})
for the top-10 recommendations.

\textbf{Accuracy}
measures the proportion of samples
for which the suggested completion token exactly matches the single target label.

\textbf{MRR}
assesses the whole top $N$ recommended completions
and takes into account the first position 
the target is matched~\cite{radev2002evaluating}.
For a single query, the reciprocal rank is
$\frac{1}{rank}$
where $rank$ is the position of the highest-ranked answer
$(1, 2, 3, ..., N$ for $N$ answers).
If no correct answer exists in top-$N$,
then the reciprocal rank is $0$.
For multiple queries $Q$,
the MRR is the mean of the $Q$ reciprocal ranks.

\subsubsection{Statement Level Prediction (SLP)}
\begin{figure}[tb!]
	\centerline{\includegraphics[width=0.6\linewidth]{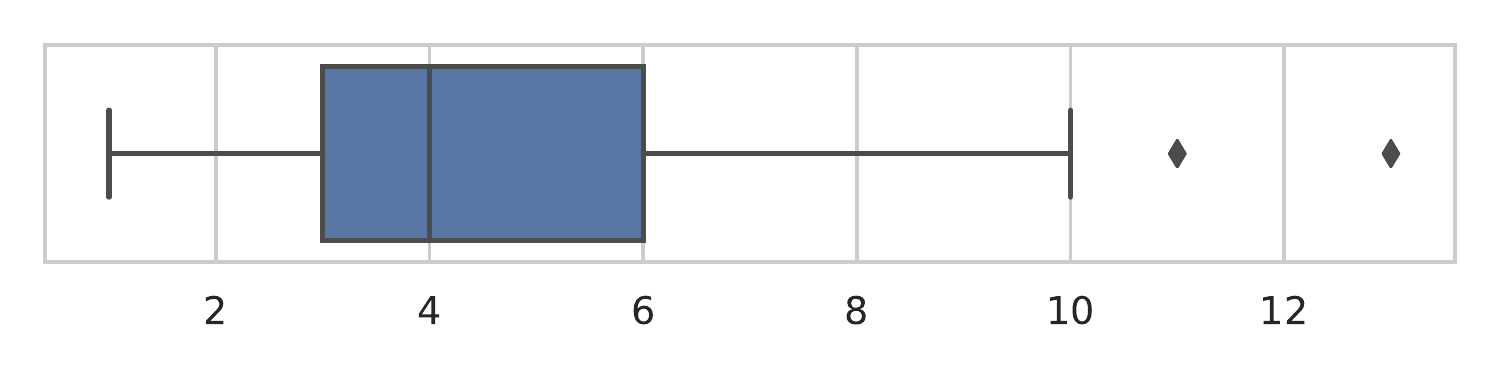}}
	\caption{Length of statements in the PY117K dataset}
	\label{fig:stats_tokens_completion}
\end{figure}

The SLP task
assesses a model's ability to complete
statements with up to $n$ tokens.
The boxplot in
Figure~\ref{fig:stats_tokens_completion}
shows the distribution of number of tokens
for completions
in the evaluation dataset ($PY117K$).
In our datasets,
statements are $4.2$ tokens long on average (median: $4$, maximum: $13$).
To provide a comprehensive view,
we evaluate the performance of the models
when predicting next-$n$ tokens with $n \in [2,3,\ldots, 8]$.

\paragraph{Evaluation Metrics:}\label{sec:SLP_metrics}
On absence of code-specific metrics,
we use two metrics commonly-used for automatic evaluation of
text generation,
namely Metric for Evaluation of Translation with Explicit ORdering (METEOR)~\cite{lavie2004significance}
and Recall-Oriented Understudy for Gisting Evaluation (ROUGE-L)~\cite{lin2004rouge}.

\textbf{ROUGE}:
ROUGE-N refers to overlapping n-grams.
ROUGE-L, one of the variations of the ROUGE metric,
counts longest matching sequence of words
using the Longest Common Subsequence algorithm.
It considers sentence-level structure similarity
and automatically identifies the longest co-occurring chain 
of in sequence n-grams.
Thus, it does not require consecutive matches
but in-sequence matches that reflect sentence-level word order.

\textbf{METEOR}
 is based on the term-to-term mapping of the generated code
with its corresponding reference code.
It focuses mainly on recall.
Lavie et al.~\cite{lavie2004significance} showed
metrics based on recall consistently achieve
higher correlation with user preferences
than those based on precision alone.

\subsection{Datasets}\label{sec:datasets}
We use two Python datasets for training and evaluation:
\begin{itemize}
    \item The ETH $150$K Python dataset~\cite{raychev2016probabilistic}
for compatibility with previous work.
The authors collected Python programs from GitHub repositories
and removed duplicate files, project forks, files that do not parse and have more
than $30$K nodes in their ASTs.
They also removed obfuscated files
and only used repositories with permissive licenses
including MIT, BSD, and Apache.
    \item The \cf{} dataset, which was collected
    by querying GHTorrent~\cite{gousios2012ghtorrent} for all non-forked Python
    repositories with more than $20$ stars ($58$k repositories).
\end{itemize}

After deduplication, using the method 
proposed by Allamanis~\cite{allamanis2019adverse},
we ended up with two versions of the original datasets, $PY117K$ and $PY1690K$
for the ETH and \cf{} datasets, respectively.
Note that $PY1690K$ and $PY117K$  do not have any common files.
Table~\ref{tab:dataset_info}
presents an overview of the contents of the datasets.

We use $PY1690K$ exclusively for pre-training our LM.
We then use $90\%$ of $PY117K$
for fine-tuning the model on the tasks presented in Section~\ref{sec:eval_tasks},
and finally the last $10\%$ of $PY117K$ for evaluation.
For the baselines,
we concatenate $PY1690K$
with the same $90\%$ portion of $PY117K$ as above for training,
and evaluate on the remaining $10\%$ of $PY117K$.

\begin{table}[tb!]
\caption{Datasets used for training and evaluation}
\centering
\begin{tabular}{ccc}
\toprule
    & PY1690K & PY117K\\\midrule
    \#Repositories   & $32.7$K & $24.9$K \\
    \#Files          & $1.7$M & $117$K \\
    \#LOC            & $425$M & $29$M \\
    \#Tokens (unique) & $5.7$M & $766$K \\
    \#Types (unique)  & $103$ & $103$ \\
    \bottomrule
\label{tab:dataset_info}
\end{tabular}
\end{table}

\subsection{Implementation and Configuration}\label{sec:config}
We use Python's AST~\footnote{\url{https://docs.python.org/3/library/ast.html}},
Tokenize~\footnote{\url{https://docs.python.org/3/library/tokenize.html}},
and the
DIS~\footnote{\url{https://docs.python.org/3/library/dis.html}}
libraries in our conversion tool.
Moreover, we use the \textit{HuggingFace}~\footnote{https://huggingface.co}
library for the implementation of our GPT-2 and MTL models.
We set the learning rate to $0.00001$,
maximum sequence length to $2048$,
and trained our model for $100$ epochs.
We set the remaining parameters to default values.
Our experiments are conducted on a machine
equipped with two GeForce GTX 1080 Ti GPUs,
an Intel(R) Xeon(R) CPU E5-2690 v4 @ 2.60GHz CPU
with $14$ core processors, and $128$G RAM.

\section{Results and Discussion}\label{sec:results}
In this section, we present the results for each evaluation task,
along with an ablation study and a characterization of the models'
performance.

\subsection{Token-level Prediction (TLP)}
\subsubsection{Any token prediction}
The most basic form of evaluation for an autocompletion model
is to gauge its ability
to predict the next token given some context as input.
TLP-A can provide an overview
on the ability of an autocompleter to predict,
however, it does not account for
the prior probabilities of different types
of tokens. We present this task for compatibility
with existing work, and further elaborate \cf{}'s performance
in the following tasks.
The results can be seen in Table~\ref{tab:results_TPL_A};
our model outperforms all the baselines
across all metrics.

\begin{table}
\caption{TPL-A results: Any token prediction}
\centering
\begin{small}
\begin{tabular}{cccc}
\toprule
    Approach & Venue & {Accuracy} & {MRR} \\
    \midrule
    n-gram + LSTM~\cite{hellendoorn2017deep} & (FSE, 2017) & 65.1 & 67.9 \\
    Pointer Mixture~\cite{li2017code} &  (IJCAI, 2018) & 65.8 & 70.0 \\
    OpenVocab~\cite{karampatsis2020big} &  (ICSE, 2020) & 67.2 & 69.8 \\
    T-XL + Bi-LSTM~\cite{liu2020self} & (ICPC, 2020) & 75.0 & 76.4 \\
    GPT-C~\cite{svyatkovskiy2020intellicode} & (FSE, 2020) & 79.8 & 80.0 \\
    TravTrans+~\cite{kim2020code}& (ICSE, 2021) & 78.9 & 79.4 \\\midrule
    \cf{} & Proposed & \textbf{80.6} & \textbf{81.7} \\
    \bottomrule
\label{tab:results_TPL_A}
\end{tabular}
\end{small}
\end{table}

\subsubsection{Token Type Prediction}
We investigate the performance of the models
when predicting different types of tokens, i.e., their
ability to assimilate how developers use grammar to express concepts.
Models generally struggle more with specific token types.
For instance, it is known that predicting identifiers
is harder than predicting keywords~\cite{hellendoorn2019code}.
Table~\ref{tab:results_TPL_B}
present the Accuracy and MRR results based on all token types.
As demonstrated, \cf{} outperforms the baselines
for all token types based on both metrics (except for MRR on keywords
and punctuation, where its performance is on par).
Transformer-based approaches are highly capable of
predicting specific types of tokens, namely keywords and punctuation;
effectively, this means that given enough training examples, they
can efficiently learn syntactical patterns.
Predicting identifiers and literals across all models is more challenging.
For identifiers, all models' result
range from $37\%$ to $56\%$ accuracy.
In both cases, \cf{}  maintains a non-trivial edge over the baselines,
which we attribute to the statement completion task. We believe it helps
\cf{}  to learn syntactical patterns over longer ranges.
\begin{table}
\caption{TPL-B Results: Token type predictions}
\centering
\begin{small}
\begin{tabular}{cccccccc}
\toprule
    \rotatebox{90}{Metric} & Approach
    & \rotatebox{90}{Identifier} & \rotatebox{90}{Keyword}
    & \rotatebox{90}{Punctuation} & \rotatebox{90}{Literals}
    & \rotatebox{90}{Operators} & \rotatebox{90}{All} \\
    \midrule
    & Token Percentage & 21\% & 28\% & 33\% & 5\% & 13\% & -  \\\midrule
    \multirow{7}{*}{\rotatebox{90}{Accuracy}}
    & N-gram+LSTM~\cite{hellendoorn2017deep}
    & 40.2 & 74.2 & 81.4 & 46.2 & 62.7 & 66.6 \\
    & Pointer Mixture~\cite{li2017code}
    & 37.0 & 85.3 & 80.0 & 43.9 & 62.8 & 68.4\\
    & OpenVocab~\cite{karampatsis2020big}
    & 42.3 & 89.8 & 93.4 & 54.4 & 65.0 & 76.0\\
    & T-XL + Bi-LSTM~\cite{liu2020self}
    & 47.4 & 93.1 & 92.4 & 59.4 & 68.7 & 78.4\\
    & GPT-C~\cite{svyatkovskiy2020intellicode}
    & 50.0 & 96.5 & 95.1 & 62.0 & 71.0 & 81.2\\
    & TravTrans+~\cite{kim2020code}
    & 51.1 & 95.9 & 97.0 & 59.3 & 71.3 & 81.8 \\
    \cmidrule{2-8}
    & \cf{}
    & \textbf{54.4} & \textbf{97.3} & \textbf{98.0} & \textbf{65.8} & \textbf{71.4} & \textbf{83.8} \\
    \midrule
    \multirow{7}{*}{\rotatebox{90}{MRR}}
    & N-gram+LSTM~\cite{hellendoorn2017deep}
    & 40.6 & 76.8 & 84.6 & 49.8 & 64.2 & 68.8\\
    & Pointer Mixture~\cite{li2017code}
    & 38.5 & 85.9 & 85.2 & 46.7 & 64.5 & 71.0 \\
    & OpenVocab~\cite{karampatsis2020big}
    & 43.2 & 90.3 & 96.0 & 57.0 & 67.1 & 77.6\\
    & T-XL + Bi-LSTM~\cite{liu2020self}
    & 49.8 & 96.1 & 96.6 & 61.3 & 70.0 & 81.4\\
    & GPT-C~\cite{svyatkovskiy2020intellicode}
    & 52.3 & {\bfseries 98.8} & {\bfseries 98.8} & 64.0 & 73.3 & 83.9\\
    & TravTrans+~\cite{kim2020code}
    & 53.7 & 97.1 & 98.6 & 62.2 & 73.0 & 83.6 \\
    \cmidrule{2-8}
    & \cf{}
    & \textbf{56.0} & 98.1 & 98.0 & \textbf{66.1} & \textbf{74.4} & \textbf{87.2} \\
    \bottomrule
\label{tab:results_TPL_B}
\end{tabular}
\end{small}
\vspace{-3mm}
\end{table}

\subsubsection{Leaf Node Prediction}
We compare each model's performance
in predicting different types of \textit{leaf nodes} in an AST,
e.g., function calls, variables, and attribute names.
Tables~\ref{tab:results_TLP_C}
present the Accuracy and MRR results for this task.
\cf{} is the best model in both accuracy, and, especially, MRR.
This means that its name predictions, which is arguably the most important
feature for an autocompleter, 
are 2 out of 3 times correct and have a high
probability ($> 70\%$) of being included in the top suggestions.
\begin{table}
\caption{TLP-C results: Leaf node prediction}
\centering
\begin{small}
\begin{tabular}{ccccccc}
\toprule
    \rotatebox{90}{Metric} & Approach
    & \rotatebox{90}{\parbox{1cm}{Attribute\\Access}}
    & \rotatebox{90}{Names}
    & \rotatebox{90}{\parbox{1cm}{Function\\names}}
    & \rotatebox{90}{\parbox{1cm}{Numeric\\constant}}
    & \rotatebox{90}{All}\\
    \midrule
    & Token Percentage  & 32\% & 13\% & 33\% & 22\% &  \\\midrule
    \multirow{7}{*}{\rotatebox{90}{Accuracy}}
    & N-gram + LSTM~\cite{hellendoorn2017deep}& 56.3 & 61.8 & 63.5 & 45.1 & 56.9\\
    & Pointer Mixture~\cite{li2017code}    & 53.5 & 62.0 & 59.8 & 42.0 & 54.2 \\
    & OpenVocab~\cite{karampatsis2020big}  & 59.8 & 63.7 & 66.2 & 51.7 & 60.6 \\
    & T-XL + Bi-LSTM~\cite{liu2020self}    & 59.9 & 58.1 & 62.8 & 54.8 & 59.5 \\
    & GPT-C~\cite{svyatkovskiy2020intellicode}& 60.0 & 59.9 & 64.0 & {\bfseries 56.0} & 60.4 \\
    & TravTrans+~\cite{kim2020code}        & 60.2 & 65.4 & 68.3 & 52.7 & 61.7 \\
    \cmidrule{2-7}
    & \cf{}  & \textbf{64.0} & \textbf{67.3} & \textbf{72.2} & 53.1 & \textbf{66.3} \\
    \midrule
    \multirow{7}{*}{\rotatebox{90}{MRR}}
    & N-gram + LSTM~\cite{hellendoorn2017deep}& 57.9 & 64.7 & 65.2 & 47.5 & 58.9\\
    & Pointer Mixture~\cite{li2017code}    & 57.1 & 59.0 & 60.2 & 43.1 & 55.3 \\
    & OpenVocab~\cite{karampatsis2020big}  & 61.2 & 64.8 & 70.1 & 51.7 & 62.5 \\
    & T-XL + Bi-LSTM~\cite{liu2020self}    & 61.9 & 65.3 & 69.9 & 55.3 & 63.5 \\
    & GPT-C~\cite{svyatkovskiy2020intellicode}& 63.4 & 62.9 & 66.5 & \textbf{57.2} & 63.0 \\
    & TravTrans+~\cite{kim2020code}        & 62.8 & 65.4 & 70.0 & 55.2 & 63.8 \\
    \cmidrule{2-7}
    & \cf{}  & \textbf{72.0} & \textbf{69.7} & \textbf{76.9} & 56.0 & \textbf{69.5} \\
    \bottomrule
\label{tab:results_TLP_C}
\end{tabular}
\end{small}
\end{table}

\subsubsection{Cardinal Point Completion}\label{sec:cardinal-completion}
In Table~\ref{tab:results_TPL_D}, we report the performance of models
when predicting at cardinal points
(described in Section~\ref{sec:eval_tasks}).
As indicated,
\cf{}  outperforms all the baselines.
Consequently, it is more capable of presenting correct recommendations
at points where autocompletion is more likely to be triggered.

\begin{table}
\caption{TPL-D Results: Cardinal Points Completion}
\begin{small}
\centering
\begin{tabular}{ccc}
\toprule
    Approach & {Accuracy} & {MRR} \\
    \midrule
    N-gram + LSTM~\cite{hellendoorn2017deep}
    & 49.0 & 52.3   \\
    Pointer Mixture~\cite{li2017code}
    & 51.3 & 52.4  \\
    OpenVocab~\cite{karampatsis2020big}
    & 52.2 & 53.5  \\
    T-XL + Bi-LSTM~\cite{liu2020self}
    & 64.0 & 64.7  \\
    GPT-C~\cite{svyatkovskiy2020intellicode}
    & 66.1 & 67.8   \\
    TravTrans+~\cite{kim2020code}
    & 65.0 & 66.2 \\\midrule
    \cf{}
    & \textbf{70.0} & \textbf{70.9}  \\
    \bottomrule
\label{tab:results_TPL_D}
\end{tabular}
\end{small}
\end{table}

\subsection{Statement-Level Prediction (SLP)}
\begin{figure*}
    \centering
    \subfigure
        {\includegraphics[width=0.4\linewidth]{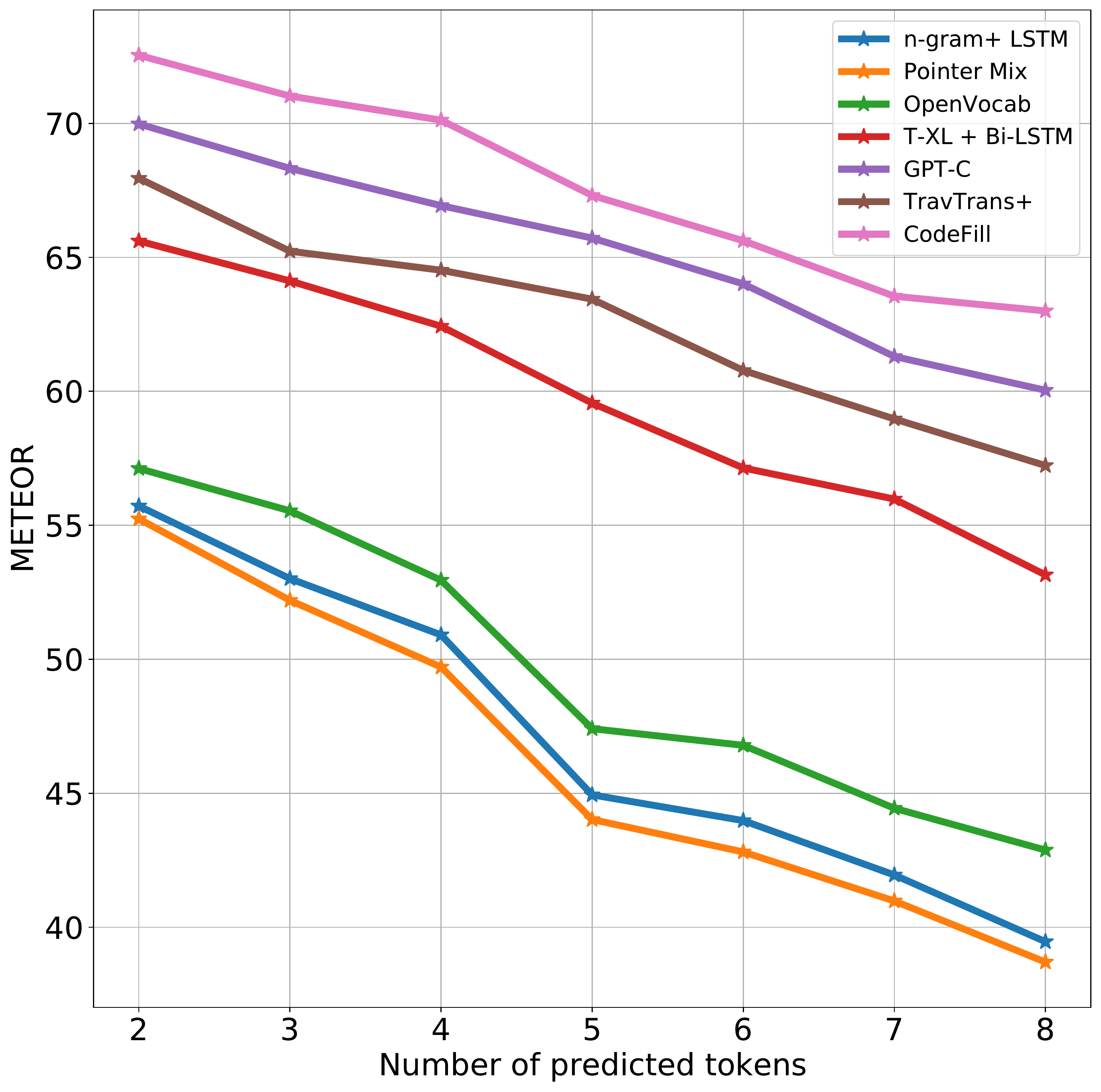}}
    \subfigure
        {\includegraphics[width=0.4\linewidth]{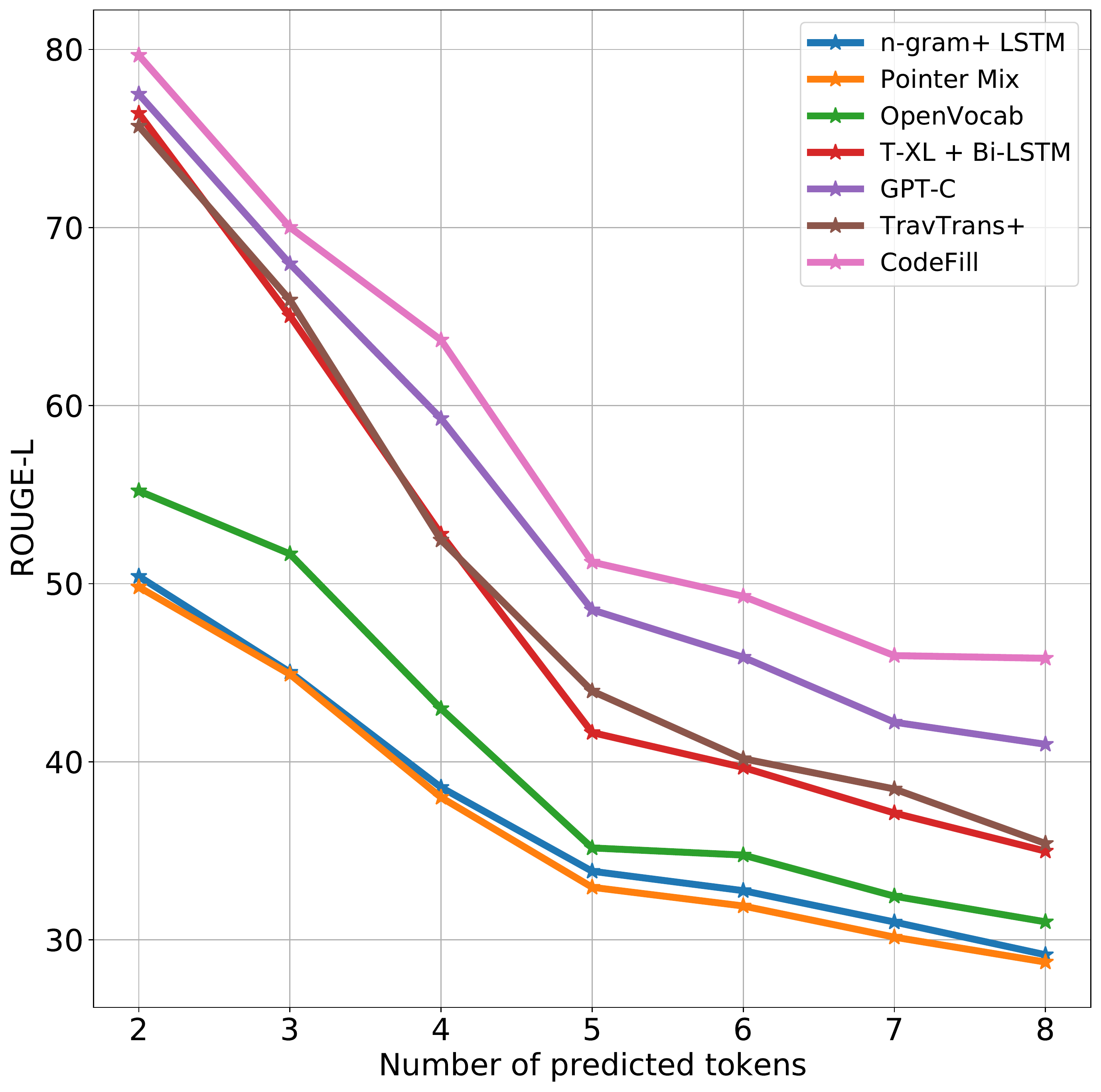}}
    \caption{Results for the SLP task}
    \label{fig:statement}
\end{figure*}

We report the results for autocompleting code statements,
by predicting the remaining $n$ tokens at a given statement position
(with $n$ ranging between $2$ and $8$).
Figure~\ref{fig:statement}
presents the results of this experiment
based on the achieved METEOR and ROUGE-L scores.
All Transformer-based models~\cite{liu2020self,svyatkovskiy2020intellicode,kim2020code},
are consistently more capable than the three baseline approaches.
\cf{} improves over
all competitors.
The margin grows wider
as the number of tokens required
to complete statements increase
(especially in the ROUGE-L case).
This result highlights the merits of our statement completion task.
In turn, this can help developers code faster
by reducing the number of required keystrokes;
the experience of using statement completion should be reminiscent 
of text line completion in popular
online email or document editors.
Statistically, \emph{more than 2 out of 3 statement completions of 4 or fewer tokens will be correct.}

\subsection{Ablation Study}
We perform an ablation study to examine the impact
of different components of \cf{} .
Table~\ref{tab:results_ablation}
presents the results of this study.
We include the performance of a vanilla GPT-2 model
to show the importance of employing the MTL approach
to jointly train models on different representations of source code.
The results show that employing the MTL technique to
train the models jointly on multiple tasks
indeed helps the model learn better.
Next, we conduct experiments to compare
hard-parameter and soft-parameter models with the two-task MTL model.
It is worth mentioning
that for the hard-parameter sharing variation,
we need to input a unified representation to the models.
Thus, we concatenate the type and value of each token
as $x_i = [t_i, v_i]$
and then feed the vectors of this concatenated representation to the MTL model.
The results indicate that
the soft-parameter sharing works better in our case.
This is probably because this setting
allows each task to have its own model and parameters
and then regularizes the distance between them
to encourage the parameters to be similar.
Finally, to verify whether adding information regarding statements helps,
we investigate the effect of adding the third task,
\textit{statement completion}.
The results demonstrate that training
on two different granularity (single-token and statement)
also helps them learn better.
To conclude,
each component of the proposed model adds to its value.
Although the training time increases,
it can be argued that training time is a one-time cost,
and can be significantly reduced with parallel training on 
multiple GPUs.
\begin{table}
\caption{Effectiveness of Different Components of the Model}
\centering
\begin{small}
\begin{tabular}{cccccc}
\toprule
    Approach & Tasks  & Train Time  & Accuracy & MRR\\
    \midrule
    GPT-2  &   Value   & 12h & 77.7 & 78.2\\
    MTL HP & Value, Type & 17h & 78.3 & 79.6\\
    MTL SP & Value, Type  & 19h & 78.9 & 79.5\\\midrule
    MTL SP & Value, Type, Statement & 24h & \textbf{80.6} & \textbf{81.7} \\
    \bottomrule
\label{tab:results_ablation}
\end{tabular}
\end{small}
\end{table}

\subsection{Runtime Characteristics}
An important aspect of ML-based autocompletion tools is
their prediction latency. A very accurate model that takes
$1$ second per prediction will not be very useful in practice
as it will interfere with the developer's workflow.
As Table~\ref{tab:complexity}, all models feature an average
latency of less than $100$ milliseconds, which is considered
the golden standard in the industry.

Moreover, the model size and number of parameters
are important practical aspects
that affect a model's deployment; if the model is too big,
it will need to be deployed centrally and clients should
connect to the model server over a network connection (which
may affect latency negatively), otherwise, it could be distributed
to the clients.
As Table~\ref{tab:complexity} shows, 
\cf{}'s number of parameters is more than other baselines 
due to our architecture specification. 
However, the size of all Transformer-based models 
makes them impractical 
for distribution to clients,
necessitating centralized deployments.
\begin{table}
\caption{Runtime Characteristics}
\centering
\begin{tabular}{cccc}
\toprule
    Approach & \parbox{2cm}{Train Time (hr)} & \parbox{1.7cm}{Latency (ms)} & \#Params\\
    \midrule
    n-gram + LSTM~\cite{hellendoorn2017deep} & 23 & 75 & 168M\\ 
    Pointer Mixture~\cite{li2017code} & 18 & 62 & 177M\\ 
    OpenVocab~\cite{karampatsis2020big} & 21 & 61 & 145M\\ 
    T-XL + Bi-LSTM~\cite{liu2020self} & 24 & 79 & 173M\\ 
    GPT-C~\cite{svyatkovskiy2020intellicode} & 23 & 74 & 125M\\ 
    TravTrans+~\cite{kim2020code} & 15 & 53 & 119M \\
    \midrule
    \cf{}  & 24 & 73 & 258M\\
    \bottomrule
\label{tab:complexity}
\end{tabular}
\end{table}

\section{Contributions and Implications}
Autocompletion is a popular research area,
however, the existing challenges
leave substantial margin for improvement,
particularly for recommending identifiers
or completing longer sequences~\mbox{\cite{hellendoorn2019code}}.
In this study, \cf{}  learns from sequences of
both token types and token names simultaneously using MTL.
The contribution of this work is twofold;

\textbf{\textit{Technical novelty}}:
Similar to the
state-of-the-art~\mbox{\cite{kim2020code,svyatkovskiy2020intellicode}},
we use transformers
for learning a name-based sequencing model,
and similar to the studies 
by Liu et al.~\mbox{\cite{liu2020self,liu2020multi}},
we use the MTL technique
to condition our models under different tasks.
However, IntelliCodeCompose~\mbox{\cite{svyatkovskiy2020intellicode}}
treats code as natural text,
neglecting the rich structure inherent in programs.
Moreover they focus on multi-lingual LMs.
TravTrans+~\mbox{\cite{kim2020code}} uses serialized ASTs
in an attempt to learn from structure,
however, we show that our novel transformation,
which we designed so that it is closer to
how developers treat source code structure, outperforms TravTrans+.
\cf{} also learns from our novel statement completion task
to consider longer contexts.
Both \mbox{\autoref{fig:statement}}
and \mbox{\autoref{tab:results_ablation}}
show that this technique improves the model,
probably by helping it better utilize completion context.
The combination of the above demonstrably
results in higher evaluation scores and better recommendations.

\textbf{\textit{Evaluation}}:
We propose two novel evaluation tasks,
\textit{cardinal point}, and \textit{statement completion},
to address deficiencies in current autocompletion evaluation setups.
We also collect, pre-process, deduplicate,
and share an large Python dataset,
consisting of practically all Python code on GitHub.

\section{Threats to the Validity}\label{sec:threats}
\textbf{Threats to internal validity}:
These include the threats
pertaining to the parameters affecting the performance of the model.
Another threat in this section relates to the errors in
the implementation of the baselines.
For all of these approaches,
we have used the replication packages provided by these studies.

\textbf{Threats to external validity}:
These threats relate to the quality of the datasets
we used and the generalizability of the results.
We used two Python datasets;
PY117K is a benchmark dataset ~\cite{raychev2016probabilistic} 
frequently used in the 
literature~\cite{li2017code,liu2020self,karampatsis2020big,kim2020code}.
PY1690K, our second dataset, is ten times larger
with approximately $1.7M$ program files.
More data can lead to more generalizable results.
Furthermore, as Allamanis.~\cite{allamanis2019adverse} suggests,
we have de-duplicated both datasets to avoid biasing the models.
All of the programs in both datasets are collected from
open-source GitHub repositories.
However, further studies are
needed to validate and generalize our findings
to other programming
languages.

\textbf{Threats to construct validity}:
These relate to the suitability of the evaluation setting and metrics.
In this work, we have tried to incorporate diverse evaluation measures.
For the TLP task,
we have used standard evaluation metrics,
namely Accuracy and MRR
in the top-one and top-ten recommendations
which are both frequently used in the literature~\cite{li2017code,liu2020self,karampatsis2020big,kim2020code}.
Furthermore, we use ROUGE-L and METEOR scores for evaluation
in the SLP task as used in previous studies
on source sequence of code generation, summarization, and translation~\cite{svyatkovskiy2020intellicode,aghamohammadi2020generating}.

\section{Conclusion and Future Work}\label{sec:conclusion}
Unlike natural language text, source code is more structured, 
its grammar is more well defined but its vocabulary is orders
of magnitude bigger.
Consequently, NLP-based models and corresponding evaluation methods
need to be adapted to the particular case of source code.

In this work, we proposed \cf{},
a Transformer-based generative LM for source code
pre-trained on three tasks closely relevant to programming.
Given a context of tokens (and their types), 
\cf{} is trained to predict
(1) the type of the next token,
(2) its value,
and (3) the values of up to $n$ next tokens.
We employ the MTL approach
to jointly train \cf{} on the above tasks.
We also propose 2 novel evaluation tasks, 
cardinal point prediction and statement-level multi-token prediction,
which we argue that they better represent how autocompletion systems
are used in practice.
We extensively evaluate \cf{} 
against six baselines on both tasks.
Our results indicate that \cf{} outperforms 
all the baselines in all scenarios,
achieving state of the art scores on both accuracy (80.6\%) 
and MRR (81.7\%) in the basic token-level prediction task.
Moreover, we show that \cf{} also learns to autocomplete statements
of up to 4 tokens with over 70\% accuracy, 
a significant improvement over the baselines, 
making it practical to offer statement completions as an IDE feature.

In the future, we plan to incorporate more domain specific
knowledge on aspects of training and evaluating a
training ML models.
For instance, one can limit the context fed to the model
based on the programming language
to better incorporate related information of functions 
and nested scopes in a piece of code.
We also plan to further investigate statement completion,
including better metrics for its evaluation.

\begin{acks}
This work has received funding from
the European Union’s Horizon 2020 research and innovation programme
under grant number 825328 (FASTEN project),
and also the NWO MIPL project, 
grant number 628.008.003.
\end{acks}
\bibliographystyle{ACM-Reference-Format}
\bibliography{main}


\begin{thebibliography}{56}


\ifx \showCODEN    \undefined \def \showCODEN     #1{\unskip}     \fi
\ifx \showDOI      \undefined \def \showDOI       #1{#1}\fi
\ifx \showISBNx    \undefined \def \showISBNx     #1{\unskip}     \fi
\ifx \showISBNxiii \undefined \def \showISBNxiii  #1{\unskip}     \fi
\ifx \showISSN     \undefined \def \showISSN      #1{\unskip}     \fi
\ifx \showLCCN     \undefined \def \showLCCN      #1{\unskip}     \fi
\ifx \shownote     \undefined \def \shownote      #1{#1}          \fi
\ifx \showarticletitle \undefined \def \showarticletitle #1{#1}   \fi
\ifx \showURL      \undefined \def \showURL       {\relax}        \fi
\providecommand\bibfield[2]{#2}
\providecommand\bibinfo[2]{#2}
\providecommand\natexlab[1]{#1}
\providecommand\showeprint[2][]{arXiv:#2}

\bibitem[\protect\citeauthoryear{Aghamohammadi, Izadi, and
  Heydarnoori}{Aghamohammadi et~al\mbox{.}}{2020}]%
        {aghamohammadi2020generating}
\bibfield{author}{\bibinfo{person}{Alireza Aghamohammadi},
  \bibinfo{person}{Maliheh Izadi}, {and} \bibinfo{person}{Abbas Heydarnoori}.}
  \bibinfo{year}{2020}\natexlab{}.
\newblock \showarticletitle{Generating summaries for methods of event-driven
  programs: An Android case study}.
\newblock \bibinfo{journal}{\emph{Journal of Systems and Software}}
  \bibinfo{volume}{170} (\bibinfo{year}{2020}), \bibinfo{pages}{110800}.
\newblock
\urldef\tempurl%
\url{https://doi.org/10.1016/j.jss.2020.110800}
\showDOI{\tempurl}


\bibitem[\protect\citeauthoryear{Allamanis}{Allamanis}{2019}]%
        {allamanis2019adverse}
\bibfield{author}{\bibinfo{person}{Miltiadis Allamanis}.}
  \bibinfo{year}{2019}\natexlab{}.
\newblock \showarticletitle{The adverse effects of code duplication in machine
  learning models of code}. In \bibinfo{booktitle}{\emph{Proceedings of the
  2019 ACM SIGPLAN International Symposium on New Ideas, New Paradigms, and
  Reflections on Programming and Software}}. \bibinfo{pages}{143--153}.
\newblock


\bibitem[\protect\citeauthoryear{Allamanis, Barr, Devanbu, and
  Sutton}{Allamanis et~al\mbox{.}}{2018}]%
        {allamanis2018survey}
\bibfield{author}{\bibinfo{person}{Miltiadis Allamanis},
  \bibinfo{person}{Earl~T Barr}, \bibinfo{person}{Premkumar Devanbu}, {and}
  \bibinfo{person}{Charles Sutton}.} \bibinfo{year}{2018}\natexlab{}.
\newblock \showarticletitle{A survey of machine learning for big code and
  naturalness}.
\newblock \bibinfo{journal}{\emph{ACM Computing Surveys (CSUR)}}
  \bibinfo{volume}{51}, \bibinfo{number}{4} (\bibinfo{year}{2018}),
  \bibinfo{pages}{1--37}.
\newblock


\bibitem[\protect\citeauthoryear{Allamanis and Sutton}{Allamanis and
  Sutton}{2014}]%
        {allamanis2014mining}
\bibfield{author}{\bibinfo{person}{Miltiadis Allamanis} {and}
  \bibinfo{person}{Charles Sutton}.} \bibinfo{year}{2014}\natexlab{}.
\newblock \showarticletitle{Mining idioms from source code}. In
  \bibinfo{booktitle}{\emph{Proceedings of the 22nd ACM SIGSOFT International
  Symposium on Foundations of Software Engineering}}.
  \bibinfo{pages}{472--483}.
\newblock


\bibitem[\protect\citeauthoryear{Amann, Proksch, Nadi, and Mezini}{Amann
  et~al\mbox{.}}{2016}]%
        {amann2016study}
\bibfield{author}{\bibinfo{person}{Sven Amann}, \bibinfo{person}{Sebastian
  Proksch}, \bibinfo{person}{Sarah Nadi}, {and} \bibinfo{person}{Mira Mezini}.}
  \bibinfo{year}{2016}\natexlab{}.
\newblock \showarticletitle{A study of visual studio usage in practice}. In
  \bibinfo{booktitle}{\emph{2016 IEEE 23rd International Conference on Software
  Analysis, Evolution, and Reengineering (SANER)}}, Vol.~\bibinfo{volume}{1}.
  IEEE, \bibinfo{pages}{124--134}.
\newblock


\bibitem[\protect\citeauthoryear{Aye and Kaiser}{Aye and Kaiser}{2020}]%
        {aye2020sequence}
\bibfield{author}{\bibinfo{person}{Gareth~Ari Aye} {and}
  \bibinfo{person}{Gail~E Kaiser}.} \bibinfo{year}{2020}\natexlab{}.
\newblock \showarticletitle{Sequence model design for code completion in the
  modern IDE}.
\newblock \bibinfo{journal}{\emph{arXiv preprint arXiv:2004.05249}}
  (\bibinfo{year}{2020}).
\newblock


\bibitem[\protect\citeauthoryear{Aye, Kim, and Li}{Aye et~al\mbox{.}}{2021}]%
        {aye2021learning}
\bibfield{author}{\bibinfo{person}{Gareth~Ari Aye}, \bibinfo{person}{Seohyun
  Kim}, {and} \bibinfo{person}{Hongyu Li}.} \bibinfo{year}{2021}\natexlab{}.
\newblock \showarticletitle{Learning autocompletion from real-world datasets}.
  In \bibinfo{booktitle}{\emph{2021 IEEE/ACM 43rd International Conference on
  Software Engineering: Software Engineering in Practice (ICSE-SEIP)}}. IEEE,
  \bibinfo{pages}{131--139}.
\newblock


\bibitem[\protect\citeauthoryear{Bielik, Raychev, and Vechev}{Bielik
  et~al\mbox{.}}{2016}]%
        {bielik2016phog}
\bibfield{author}{\bibinfo{person}{Pavol Bielik}, \bibinfo{person}{Veselin
  Raychev}, {and} \bibinfo{person}{Martin Vechev}.}
  \bibinfo{year}{2016}\natexlab{}.
\newblock \showarticletitle{PHOG: probabilistic model for code}. In
  \bibinfo{booktitle}{\emph{International Conference on Machine Learning}}.
  \bibinfo{pages}{2933--2942}.
\newblock


\bibitem[\protect\citeauthoryear{Bruch, Monperrus, and Mezini}{Bruch
  et~al\mbox{.}}{2009}]%
        {bruch2009learning}
\bibfield{author}{\bibinfo{person}{Marcel Bruch}, \bibinfo{person}{Martin
  Monperrus}, {and} \bibinfo{person}{Mira Mezini}.}
  \bibinfo{year}{2009}\natexlab{}.
\newblock \showarticletitle{Learning from examples to improve code completion
  systems}. In \bibinfo{booktitle}{\emph{Proceedings of the 7th joint meeting
  of the European software engineering conference and the ACM SIGSOFT symposium
  on the foundations of software engineering}}. \bibinfo{pages}{213--222}.
\newblock


\bibitem[\protect\citeauthoryear{Budzianowski and Vuli{\'c}}{Budzianowski and
  Vuli{\'c}}{2019}]%
        {budzianowski2019hello}
\bibfield{author}{\bibinfo{person}{Pawe{\l} Budzianowski} {and}
  \bibinfo{person}{Ivan Vuli{\'c}}.} \bibinfo{year}{2019}\natexlab{}.
\newblock \showarticletitle{Hello, It’s GPT-2-How Can I Help You? Towards the
  Use of Pretrained Language Models for Task-Oriented Dialogue Systems}. In
  \bibinfo{booktitle}{\emph{Proceedings of the 3rd Workshop on Neural
  Generation and Translation}}. \bibinfo{pages}{15--22}.
\newblock


\bibitem[\protect\citeauthoryear{Caruana}{Caruana}{1997}]%
        {caruana1997multitask}
\bibfield{author}{\bibinfo{person}{Rich Caruana}.}
  \bibinfo{year}{1997}\natexlab{}.
\newblock \showarticletitle{Multitask learning}.
\newblock \bibinfo{journal}{\emph{Machine learning}} \bibinfo{volume}{28},
  \bibinfo{number}{1} (\bibinfo{year}{1997}), \bibinfo{pages}{41--75}.
\newblock


\bibitem[\protect\citeauthoryear{Dash, Allamanis, and Barr}{Dash
  et~al\mbox{.}}{2018}]%
        {santanu2018refinym}
\bibfield{author}{\bibinfo{person}{Santanu~Kumar Dash},
  \bibinfo{person}{Miltiadis Allamanis}, {and} \bibinfo{person}{Earl~T. Barr}.}
  \bibinfo{year}{2018}\natexlab{}.
\newblock \showarticletitle{RefiNym: Using Names to Refine Types}. In
  \bibinfo{booktitle}{\emph{Proceedings of the 2018 26th ACM Joint Meeting on
  European Software Engineering Conference and Symposium on the Foundations of
  Software Engineering}} (Lake Buena Vista, FL, USA)
  \emph{(\bibinfo{series}{ESEC/FSE 2018})}. \bibinfo{publisher}{Association for
  Computing Machinery}, \bibinfo{address}{New York, NY, USA},
  \bibinfo{pages}{107–117}.
\newblock
\showISBNx{9781450355735}
\urldef\tempurl%
\url{https://doi.org/10.1145/3236024.3236042}
\showDOI{\tempurl}


\bibitem[\protect\citeauthoryear{Devlin, Chang, Lee, and Toutanova}{Devlin
  et~al\mbox{.}}{2018}]%
        {devlin2018bert}
\bibfield{author}{\bibinfo{person}{Jacob Devlin}, \bibinfo{person}{Ming-Wei
  Chang}, \bibinfo{person}{Kenton Lee}, {and} \bibinfo{person}{Kristina
  Toutanova}.} \bibinfo{year}{2018}\natexlab{}.
\newblock \showarticletitle{Bert: Pre-training of deep bidirectional
  transformers for language understanding}.
\newblock \bibinfo{journal}{\emph{arXiv preprint arXiv:1810.04805}}
  (\bibinfo{year}{2018}).
\newblock


\bibitem[\protect\citeauthoryear{Goldberg}{Goldberg}{2017}]%
        {goldberg2017neural}
\bibfield{author}{\bibinfo{person}{Yoav Goldberg}.}
  \bibinfo{year}{2017}\natexlab{}.
\newblock \showarticletitle{Neural network methods for natural language
  processing}.
\newblock \bibinfo{journal}{\emph{Synthesis lectures on human language
  technologies}} \bibinfo{volume}{10}, \bibinfo{number}{1}
  (\bibinfo{year}{2017}), \bibinfo{pages}{1--309}.
\newblock


\bibitem[\protect\citeauthoryear{Gousios and Spinellis}{Gousios and
  Spinellis}{2012}]%
        {gousios2012ghtorrent}
\bibfield{author}{\bibinfo{person}{Georgios Gousios} {and}
  \bibinfo{person}{Diomidis Spinellis}.} \bibinfo{year}{2012}\natexlab{}.
\newblock \showarticletitle{{GHT}orrent: {G}it{H}ub's Data from a Firehose}. In
  \bibinfo{booktitle}{\emph{{MSR} '12: Proceedings of the 9th Working
  Conference on Mining Software Repositories}} (Zurich, Switzerland),
  \bibfield{editor}{\bibinfo{person}{Michael~W. Godfrey} {and}
  \bibinfo{person}{Jim Whitehead}} (Eds.). \bibinfo{publisher}{IEEE},
  \bibinfo{pages}{12--21}.
\newblock
\showISSN{2160-1852}
\urldef\tempurl%
\url{https://doi.org/10.1109/MSR.2012.6224294}
\showDOI{\tempurl}


\bibitem[\protect\citeauthoryear{Ham, Lee, Jang, and Kim}{Ham
  et~al\mbox{.}}{2020}]%
        {ham2020end}
\bibfield{author}{\bibinfo{person}{Donghoon Ham}, \bibinfo{person}{Jeong-Gwan
  Lee}, \bibinfo{person}{Youngsoo Jang}, {and} \bibinfo{person}{Kee-Eung Kim}.}
  \bibinfo{year}{2020}\natexlab{}.
\newblock \showarticletitle{End-to-end neural pipeline for goal-oriented
  dialogue systems using GPT-2}. In \bibinfo{booktitle}{\emph{Proceedings of
  the 58th Annual Meeting of the Association for Computational Linguistics}}.
  \bibinfo{pages}{583--592}.
\newblock


\bibitem[\protect\citeauthoryear{Hellendoorn and Devanbu}{Hellendoorn and
  Devanbu}{2017}]%
        {hellendoorn2017deep}
\bibfield{author}{\bibinfo{person}{Vincent~J Hellendoorn} {and}
  \bibinfo{person}{Premkumar Devanbu}.} \bibinfo{year}{2017}\natexlab{}.
\newblock \showarticletitle{Are deep neural networks the best choice for
  modeling source code?}. In \bibinfo{booktitle}{\emph{Proceedings of the 2017
  11th Joint Meeting on Foundations of Software Engineering}}.
  \bibinfo{pages}{763--773}.
\newblock


\bibitem[\protect\citeauthoryear{Hellendoorn, Proksch, Gall, and
  Bacchelli}{Hellendoorn et~al\mbox{.}}{2019}]%
        {hellendoorn2019code}
\bibfield{author}{\bibinfo{person}{Vincent~J Hellendoorn},
  \bibinfo{person}{Sebastian Proksch}, \bibinfo{person}{Harald~C Gall}, {and}
  \bibinfo{person}{Alberto Bacchelli}.} \bibinfo{year}{2019}\natexlab{}.
\newblock \showarticletitle{When code completion fails: A case study on
  real-world completions}. In \bibinfo{booktitle}{\emph{2019 IEEE/ACM 41st
  International Conference on Software Engineering (ICSE)}}. IEEE,
  \bibinfo{pages}{960--970}.
\newblock


\bibitem[\protect\citeauthoryear{Hindle, Barr, Su, Gabel, and Devanbu}{Hindle
  et~al\mbox{.}}{2012}]%
        {hindle2012naturalness}
\bibfield{author}{\bibinfo{person}{Abram Hindle}, \bibinfo{person}{Earl~T
  Barr}, \bibinfo{person}{Zhendong Su}, \bibinfo{person}{Mark Gabel}, {and}
  \bibinfo{person}{Premkumar Devanbu}.} \bibinfo{year}{2012}\natexlab{}.
\newblock \showarticletitle{On the naturalness of software}. In
  \bibinfo{booktitle}{\emph{2012 34th International Conference on Software
  Engineering (ICSE)}}. IEEE, \bibinfo{pages}{837--847}.
\newblock


\bibitem[\protect\citeauthoryear{Hou and Pletcher}{Hou and Pletcher}{2010}]%
        {hou2010towards}
\bibfield{author}{\bibinfo{person}{Daqing Hou} {and} \bibinfo{person}{David~M
  Pletcher}.} \bibinfo{year}{2010}\natexlab{}.
\newblock \showarticletitle{Towards a better code completion system by API
  grouping, filtering, and popularity-based ranking}. In
  \bibinfo{booktitle}{\emph{Proceedings of the 2nd International Workshop on
  Recommendation Systems for Software Engineering}}. \bibinfo{pages}{26--30}.
\newblock


\bibitem[\protect\citeauthoryear{Izadi, Akbari, and Heydarnoori}{Izadi
  et~al\mbox{.}}{2022}]%
        {izadi2022predicting}
\bibfield{author}{\bibinfo{person}{Maliheh Izadi}, \bibinfo{person}{Kiana
  Akbari}, {and} \bibinfo{person}{Abbas Heydarnoori}.}
  \bibinfo{year}{2022}\natexlab{}.
\newblock \showarticletitle{Predicting the objective and priority of issue
  reports in software repositories}.
\newblock \bibinfo{journal}{\emph{Empirical Software Engineering}}
  \bibinfo{volume}{27}, \bibinfo{number}{2} (\bibinfo{year}{2022}),
  \bibinfo{pages}{1--37}.
\newblock
\urldef\tempurl%
\url{https://doi.org/10.1007/s10664-021-10085-3}
\showDOI{\tempurl}


\bibitem[\protect\citeauthoryear{Izadi, Heydarnoori, and Gousios}{Izadi
  et~al\mbox{.}}{2021}]%
        {izadi2021topic}
\bibfield{author}{\bibinfo{person}{Maliheh Izadi}, \bibinfo{person}{Abbas
  Heydarnoori}, {and} \bibinfo{person}{Georgios Gousios}.}
  \bibinfo{year}{2021}\natexlab{}.
\newblock \showarticletitle{Topic recommendation for software repositories
  using multi-label classification algorithms}.
\newblock \bibinfo{journal}{\emph{Empirical Software Engineering}}
  \bibinfo{volume}{26}, \bibinfo{number}{5} (\bibinfo{year}{2021}),
  \bibinfo{pages}{1--33}.
\newblock
\urldef\tempurl%
\url{https://doi.org/10.1007/s10664-021-09976-2}
\showDOI{\tempurl}


\bibitem[\protect\citeauthoryear{Jin and Servant}{Jin and Servant}{2018}]%
        {xianhao2018the}
\bibfield{author}{\bibinfo{person}{Xianhao Jin} {and}
  \bibinfo{person}{Francisco Servant}.} \bibinfo{year}{2018}\natexlab{}.
\newblock \showarticletitle{The Hidden Cost of Code Completion: Understanding
  the Impact of the Recommendation-List Length on Its Efficiency}. In
  \bibinfo{booktitle}{\emph{Proceedings of the 15th International Conference on
  Mining Software Repositories}} (Gothenburg, Sweden)
  \emph{(\bibinfo{series}{MSR '18})}. \bibinfo{publisher}{Association for
  Computing Machinery}, \bibinfo{address}{New York, NY, USA},
  \bibinfo{pages}{70–73}.
\newblock
\showISBNx{9781450357166}
\urldef\tempurl%
\url{https://doi.org/10.1145/3196398.3196474}
\showDOI{\tempurl}


\bibitem[\protect\citeauthoryear{Karampatsis, Babii, Robbes, Sutton, and
  Janes}{Karampatsis et~al\mbox{.}}{2020}]%
        {karampatsis2020big}
\bibfield{author}{\bibinfo{person}{Rafael-Michael Karampatsis},
  \bibinfo{person}{Hlib Babii}, \bibinfo{person}{Romain Robbes},
  \bibinfo{person}{Charles Sutton}, {and} \bibinfo{person}{Andrea Janes}.}
  \bibinfo{year}{2020}\natexlab{}.
\newblock \showarticletitle{Big code!= big vocabulary: Open-vocabulary models
  for source code}. In \bibinfo{booktitle}{\emph{2020 IEEE/ACM 42nd
  International Conference on Software Engineering (ICSE)}}. IEEE,
  \bibinfo{pages}{1073--1085}.
\newblock


\bibitem[\protect\citeauthoryear{Kim, Zhao, Tian, and Chandra}{Kim
  et~al\mbox{.}}{2021}]%
        {kim2020code}
\bibfield{author}{\bibinfo{person}{Seohyun Kim}, \bibinfo{person}{Jinman Zhao},
  \bibinfo{person}{Yuchi Tian}, {and} \bibinfo{person}{Satish Chandra}.}
  \bibinfo{year}{2021}\natexlab{}.
\newblock \showarticletitle{Code Prediction by Feeding Trees to Transformers}.
  In \bibinfo{booktitle}{\emph{2021 IEEE/ACM 43rd International Conference on
  Software Engineering (ICSE)}}. \bibinfo{pages}{150--162}.
\newblock
\urldef\tempurl%
\url{https://doi.org/10.1109/ICSE43902.2021.00026}
\showDOI{\tempurl}


\bibitem[\protect\citeauthoryear{Kim, Jernite, Sontag, and Rush}{Kim
  et~al\mbox{.}}{2016}]%
        {kim2016character}
\bibfield{author}{\bibinfo{person}{Yoon Kim}, \bibinfo{person}{Yacine Jernite},
  \bibinfo{person}{David Sontag}, {and} \bibinfo{person}{Alexander~M Rush}.}
  \bibinfo{year}{2016}\natexlab{}.
\newblock \showarticletitle{Character-aware neural language models}. In
  \bibinfo{booktitle}{\emph{Thirtieth AAAI conference on artificial
  intelligence}}.
\newblock


\bibitem[\protect\citeauthoryear{Lavie, Sagae, and Jayaraman}{Lavie
  et~al\mbox{.}}{2004}]%
        {lavie2004significance}
\bibfield{author}{\bibinfo{person}{Alon Lavie}, \bibinfo{person}{Kenji Sagae},
  {and} \bibinfo{person}{Shyamsundar Jayaraman}.}
  \bibinfo{year}{2004}\natexlab{}.
\newblock \showarticletitle{The significance of recall in automatic metrics for
  MT evaluation}. In \bibinfo{booktitle}{\emph{Conference of the Association
  for Machine Translation in the Americas}}. Springer,
  \bibinfo{pages}{134--143}.
\newblock


\bibitem[\protect\citeauthoryear{Lee and Hsiang}{Lee and Hsiang}{2020}]%
        {lee2020patent}
\bibfield{author}{\bibinfo{person}{Jieh-Sheng Lee} {and} \bibinfo{person}{Jieh
  Hsiang}.} \bibinfo{year}{2020}\natexlab{}.
\newblock \showarticletitle{Patent claim generation by fine-tuning OpenAI
  GPT-2}.
\newblock \bibinfo{journal}{\emph{World Patent Information}}
  \bibinfo{volume}{62} (\bibinfo{year}{2020}), \bibinfo{pages}{101983}.
\newblock


\bibitem[\protect\citeauthoryear{Li, Wang, Lyu, and King}{Li
  et~al\mbox{.}}{2017}]%
        {li2017code}
\bibfield{author}{\bibinfo{person}{Jian Li}, \bibinfo{person}{Yue Wang},
  \bibinfo{person}{Michael~R Lyu}, {and} \bibinfo{person}{Irwin King}.}
  \bibinfo{year}{2017}\natexlab{}.
\newblock \showarticletitle{Code completion with neural attention and pointer
  networks}.
\newblock \bibinfo{journal}{\emph{arXiv preprint arXiv:1711.09573}}
  (\bibinfo{year}{2017}).
\newblock


\bibitem[\protect\citeauthoryear{Lin}{Lin}{2004}]%
        {lin2004rouge}
\bibfield{author}{\bibinfo{person}{Chin-Yew Lin}.}
  \bibinfo{year}{2004}\natexlab{}.
\newblock \showarticletitle{Rouge: A package for automatic evaluation of
  summaries}. In \bibinfo{booktitle}{\emph{Text summarization branches out}}.
  \bibinfo{pages}{74--81}.
\newblock


\bibitem[\protect\citeauthoryear{Liu, Wang, Shin, Gonzalez, and Song}{Liu
  et~al\mbox{.}}{2016}]%
        {liu2016neural}
\bibfield{author}{\bibinfo{person}{Chang Liu}, \bibinfo{person}{Xin Wang},
  \bibinfo{person}{Richard Shin}, \bibinfo{person}{Joseph~E Gonzalez}, {and}
  \bibinfo{person}{Dawn Song}.} \bibinfo{year}{2016}\natexlab{}.
\newblock \showarticletitle{Neural code completion}.
\newblock  (\bibinfo{year}{2016}).
\newblock


\bibitem[\protect\citeauthoryear{Liu, Li, Wei, Xia, Fu, and Jin}{Liu
  et~al\mbox{.}}{2020a}]%
        {liu2020self}
\bibfield{author}{\bibinfo{person}{Fang Liu}, \bibinfo{person}{Ge Li},
  \bibinfo{person}{Bolin Wei}, \bibinfo{person}{Xin Xia},
  \bibinfo{person}{Zhiyi Fu}, {and} \bibinfo{person}{Zhi Jin}.}
  \bibinfo{year}{2020}\natexlab{a}.
\newblock \showarticletitle{A Self-Attentional Neural Architecture for Code
  Completion with Multi-Task Learning}. In
  \bibinfo{booktitle}{\emph{Proceedings of the 28th International Conference on
  Program Comprehension}}. \bibinfo{pages}{37--47}.
\newblock


\bibitem[\protect\citeauthoryear{Liu, Li, Zhao, and Jin}{Liu
  et~al\mbox{.}}{2020b}]%
        {liu2020multi}
\bibfield{author}{\bibinfo{person}{Fang Liu}, \bibinfo{person}{Ge Li},
  \bibinfo{person}{Yunfei Zhao}, {and} \bibinfo{person}{Zhi Jin}.}
  \bibinfo{year}{2020}\natexlab{b}.
\newblock \showarticletitle{Multi-task Learning based Pre-trained Language
  Model for Code Completion}. In \bibinfo{booktitle}{\emph{2020 35th IEEE/ACM
  International Conference on Automated Software Engineering (ASE)}}. IEEE,
  \bibinfo{pages}{473--485}.
\newblock


\bibitem[\protect\citeauthoryear{Luong, Pham, and Manning}{Luong
  et~al\mbox{.}}{2015}]%
        {luong2015effective}
\bibfield{author}{\bibinfo{person}{Minh-Thang Luong}, \bibinfo{person}{Hieu
  Pham}, {and} \bibinfo{person}{Christopher~D Manning}.}
  \bibinfo{year}{2015}\natexlab{}.
\newblock \showarticletitle{Effective approaches to attention-based neural
  machine translation}.
\newblock \bibinfo{journal}{\emph{arXiv preprint arXiv:1508.04025}}
  (\bibinfo{year}{2015}).
\newblock


\bibitem[\protect\citeauthoryear{Mikolov, Karafi{\'a}t, Burget,
  {\v{C}}ernock{\`y}, and Khudanpur}{Mikolov et~al\mbox{.}}{2010}]%
        {mikolov2010recurrent}
\bibfield{author}{\bibinfo{person}{Tom{\'a}{\v{s}} Mikolov},
  \bibinfo{person}{Martin Karafi{\'a}t}, \bibinfo{person}{Luk{\'a}{\v{s}}
  Burget}, \bibinfo{person}{Jan {\v{C}}ernock{\`y}}, {and}
  \bibinfo{person}{Sanjeev Khudanpur}.} \bibinfo{year}{2010}\natexlab{}.
\newblock \showarticletitle{Recurrent neural network based language model}. In
  \bibinfo{booktitle}{\emph{Eleventh annual conference of the international
  speech communication association}}.
\newblock


\bibitem[\protect\citeauthoryear{Nguyen, Nguyen, Li, and Wang}{Nguyen
  et~al\mbox{.}}{2019}]%
        {nguyen2019combining}
\bibfield{author}{\bibinfo{person}{Son Nguyen}, \bibinfo{person}{Tien Nguyen},
  \bibinfo{person}{Yi Li}, {and} \bibinfo{person}{Shaohua Wang}.}
  \bibinfo{year}{2019}\natexlab{}.
\newblock \showarticletitle{Combining program analysis and statistical language
  model for code statement completion}. In \bibinfo{booktitle}{\emph{2019 34th
  IEEE/ACM International Conference on Automated Software Engineering (ASE)}}.
  IEEE, \bibinfo{pages}{710--721}.
\newblock


\bibitem[\protect\citeauthoryear{Nguyen, Nguyen, Nguyen, and Nguyen}{Nguyen
  et~al\mbox{.}}{2013}]%
        {nguyen2013statistical}
\bibfield{author}{\bibinfo{person}{Tung~Thanh Nguyen},
  \bibinfo{person}{Anh~Tuan Nguyen}, \bibinfo{person}{Hoan~Anh Nguyen}, {and}
  \bibinfo{person}{Tien~N Nguyen}.} \bibinfo{year}{2013}\natexlab{}.
\newblock \showarticletitle{A statistical semantic language model for source
  code}. In \bibinfo{booktitle}{\emph{Proceedings of the 2013 9th Joint Meeting
  on Foundations of Software Engineering}}. \bibinfo{pages}{532--542}.
\newblock


\bibitem[\protect\citeauthoryear{Radev, Qi, Wu, and Fan}{Radev
  et~al\mbox{.}}{2002}]%
        {radev2002evaluating}
\bibfield{author}{\bibinfo{person}{Dragomir~R Radev}, \bibinfo{person}{Hong
  Qi}, \bibinfo{person}{Harris Wu}, {and} \bibinfo{person}{Weiguo Fan}.}
  \bibinfo{year}{2002}\natexlab{}.
\newblock \showarticletitle{Evaluating Web-based Question Answering Systems.}.
  In \bibinfo{booktitle}{\emph{LREC}}. Citeseer.
\newblock


\bibitem[\protect\citeauthoryear{Radford, Wu, Child, Luan, Amodei, and
  Sutskever}{Radford et~al\mbox{.}}{2019}]%
        {radford2019language}
\bibfield{author}{\bibinfo{person}{Alec Radford}, \bibinfo{person}{Jeff Wu},
  \bibinfo{person}{Rewon Child}, \bibinfo{person}{David Luan},
  \bibinfo{person}{Dario Amodei}, {and} \bibinfo{person}{Ilya Sutskever}.}
  \bibinfo{year}{2019}\natexlab{}.
\newblock \showarticletitle{Language Models are Unsupervised Multitask
  Learners}.
\newblock  (\bibinfo{year}{2019}).
\newblock


\bibitem[\protect\citeauthoryear{Raychev, Bielik, and Vechev}{Raychev
  et~al\mbox{.}}{2016a}]%
        {raychev2016probabilistic}
\bibfield{author}{\bibinfo{person}{Veselin Raychev}, \bibinfo{person}{Pavol
  Bielik}, {and} \bibinfo{person}{Martin Vechev}.}
  \bibinfo{year}{2016}\natexlab{a}.
\newblock \showarticletitle{Probabilistic model for code with decision trees}.
\newblock \bibinfo{journal}{\emph{ACM SIGPLAN Notices}} \bibinfo{volume}{51},
  \bibinfo{number}{10} (\bibinfo{year}{2016}), \bibinfo{pages}{731--747}.
\newblock


\bibitem[\protect\citeauthoryear{Raychev, Bielik, Vechev, and Krause}{Raychev
  et~al\mbox{.}}{2016b}]%
        {raychev2016learning}
\bibfield{author}{\bibinfo{person}{Veselin Raychev}, \bibinfo{person}{Pavol
  Bielik}, \bibinfo{person}{Martin Vechev}, {and} \bibinfo{person}{Andreas
  Krause}.} \bibinfo{year}{2016}\natexlab{b}.
\newblock \showarticletitle{Learning programs from noisy data}.
\newblock \bibinfo{journal}{\emph{ACM Sigplan Notices}} \bibinfo{volume}{51},
  \bibinfo{number}{1} (\bibinfo{year}{2016}), \bibinfo{pages}{761--774}.
\newblock


\bibitem[\protect\citeauthoryear{Robbes and Lanza}{Robbes and Lanza}{2008}]%
        {robbes2008program}
\bibfield{author}{\bibinfo{person}{Romain Robbes} {and}
  \bibinfo{person}{Michele Lanza}.} \bibinfo{year}{2008}\natexlab{}.
\newblock \showarticletitle{How program history can improve code completion}.
  In \bibinfo{booktitle}{\emph{2008 23rd IEEE/ACM International Conference on
  Automated Software Engineering}}. IEEE, \bibinfo{pages}{317--326}.
\newblock


\bibitem[\protect\citeauthoryear{Robbes and Lanza}{Robbes and Lanza}{2010}]%
        {robbes2010improving}
\bibfield{author}{\bibinfo{person}{Romain Robbes} {and}
  \bibinfo{person}{Michele Lanza}.} \bibinfo{year}{2010}\natexlab{}.
\newblock \showarticletitle{Improving code completion with program history}.
\newblock \bibinfo{journal}{\emph{Automated Software Engineering}}
  \bibinfo{volume}{17}, \bibinfo{number}{2} (\bibinfo{year}{2010}),
  \bibinfo{pages}{181--212}.
\newblock


\bibitem[\protect\citeauthoryear{Robbins and Monro}{Robbins and Monro}{1951}]%
        {robbins1951stochastic}
\bibfield{author}{\bibinfo{person}{Herbert Robbins} {and}
  \bibinfo{person}{Sutton Monro}.} \bibinfo{year}{1951}\natexlab{}.
\newblock \showarticletitle{A stochastic approximation method}.
\newblock \bibinfo{journal}{\emph{The annals of mathematical statistics}}
  (\bibinfo{year}{1951}), \bibinfo{pages}{400--407}.
\newblock


\bibitem[\protect\citeauthoryear{Ruder}{Ruder}{2017}]%
        {ruder2017overview}
\bibfield{author}{\bibinfo{person}{Sebastian Ruder}.}
  \bibinfo{year}{2017}\natexlab{}.
\newblock \showarticletitle{An overview of multi-task learning in deep neural
  networks}.
\newblock \bibinfo{journal}{\emph{arXiv preprint arXiv:1706.05098}}
  (\bibinfo{year}{2017}).
\newblock


\bibitem[\protect\citeauthoryear{Sch{\"u}tze, Manning, and
  Raghavan}{Sch{\"u}tze et~al\mbox{.}}{2008}]%
        {schutze2008introduction}
\bibfield{author}{\bibinfo{person}{Hinrich Sch{\"u}tze},
  \bibinfo{person}{Christopher~D Manning}, {and} \bibinfo{person}{Prabhakar
  Raghavan}.} \bibinfo{year}{2008}\natexlab{}.
\newblock \bibinfo{booktitle}{\emph{Introduction to information retrieval}}.
  Vol.~\bibinfo{volume}{39}.
\newblock \bibinfo{publisher}{Cambridge University Press Cambridge}.
\newblock


\bibitem[\protect\citeauthoryear{Schwenk and Gauvain}{Schwenk and
  Gauvain}{2002}]%
        {schwenk2002connectionist}
\bibfield{author}{\bibinfo{person}{Holger Schwenk} {and}
  \bibinfo{person}{Jean-Luc Gauvain}.} \bibinfo{year}{2002}\natexlab{}.
\newblock \showarticletitle{Connectionist language modeling for large
  vocabulary continuous speech recognition}. In \bibinfo{booktitle}{\emph{2002
  IEEE International Conference on Acoustics, Speech, and Signal Processing}},
  Vol.~\bibinfo{volume}{1}. IEEE, \bibinfo{pages}{I--765}.
\newblock


\bibitem[\protect\citeauthoryear{Sener and Koltun}{Sener and Koltun}{2018}]%
        {sener2018multi}
\bibfield{author}{\bibinfo{person}{Ozan Sener} {and} \bibinfo{person}{Vladlen
  Koltun}.} \bibinfo{year}{2018}\natexlab{}.
\newblock \showarticletitle{Multi-task learning as multi-objective
  optimization}.
\newblock \bibinfo{journal}{\emph{arXiv preprint arXiv:1810.04650}}
  (\bibinfo{year}{2018}).
\newblock


\bibitem[\protect\citeauthoryear{Svyatkovskiy, Deng, Fu, and
  Sundaresan}{Svyatkovskiy et~al\mbox{.}}{2020}]%
        {svyatkovskiy2020intellicode}
\bibfield{author}{\bibinfo{person}{Alexey Svyatkovskiy},
  \bibinfo{person}{Shao~Kun Deng}, \bibinfo{person}{Shengyu Fu}, {and}
  \bibinfo{person}{Neel Sundaresan}.} \bibinfo{year}{2020}\natexlab{}.
\newblock \showarticletitle{Intellicode compose: Code generation using
  transformer}. In \bibinfo{booktitle}{\emph{Proceedings of the 28th ACM Joint
  Meeting on European Software Engineering Conference and Symposium on the
  Foundations of Software Engineering}}. \bibinfo{pages}{1433--1443}.
\newblock


\bibitem[\protect\citeauthoryear{Svyatkovskiy, Lee, Hadjitofi, Riechert,
  Franco, and Allamanis}{Svyatkovskiy et~al\mbox{.}}{2021}]%
        {svyatkovskiy2021fast}
\bibfield{author}{\bibinfo{person}{Alexey Svyatkovskiy},
  \bibinfo{person}{Sebastian Lee}, \bibinfo{person}{Anna Hadjitofi},
  \bibinfo{person}{Maik Riechert}, \bibinfo{person}{Juliana~Vicente Franco},
  {and} \bibinfo{person}{Miltiadis Allamanis}.}
  \bibinfo{year}{2021}\natexlab{}.
\newblock \showarticletitle{Fast and memory-efficient neural code completion}.
  In \bibinfo{booktitle}{\emph{2021 IEEE/ACM 18th International Conference on
  Mining Software Repositories (MSR)}}. IEEE, \bibinfo{pages}{329--340}.
\newblock


\bibitem[\protect\citeauthoryear{Svyatkovskiy, Zhao, Fu, and
  Sundaresan}{Svyatkovskiy et~al\mbox{.}}{2019}]%
        {svyatkovskiy2019pythia}
\bibfield{author}{\bibinfo{person}{Alexey Svyatkovskiy}, \bibinfo{person}{Ying
  Zhao}, \bibinfo{person}{Shengyu Fu}, {and} \bibinfo{person}{Neel
  Sundaresan}.} \bibinfo{year}{2019}\natexlab{}.
\newblock \showarticletitle{Pythia: Ai-assisted code completion system}. In
  \bibinfo{booktitle}{\emph{Proceedings of the 25th ACM SIGKDD International
  Conference on Knowledge Discovery \& Data Mining}}.
  \bibinfo{pages}{2727--2735}.
\newblock


\bibitem[\protect\citeauthoryear{Tu, Su, and Devanbu}{Tu et~al\mbox{.}}{2014}]%
        {tu2014localness}
\bibfield{author}{\bibinfo{person}{Zhaopeng Tu}, \bibinfo{person}{Zhendong Su},
  {and} \bibinfo{person}{Premkumar Devanbu}.} \bibinfo{year}{2014}\natexlab{}.
\newblock \showarticletitle{On the localness of software}. In
  \bibinfo{booktitle}{\emph{Proceedings of the 22nd ACM SIGSOFT International
  Symposium on Foundations of Software Engineering}}.
  \bibinfo{pages}{269--280}.
\newblock


\bibitem[\protect\citeauthoryear{Vaswani, Shazeer, Parmar, Uszkoreit, Jones,
  Gomez, Kaiser, and Polosukhin}{Vaswani et~al\mbox{.}}{2017}]%
        {vaswani2017attention}
\bibfield{author}{\bibinfo{person}{Ashish Vaswani}, \bibinfo{person}{Noam
  Shazeer}, \bibinfo{person}{Niki Parmar}, \bibinfo{person}{Jakob Uszkoreit},
  \bibinfo{person}{Llion Jones}, \bibinfo{person}{Aidan~N Gomez},
  \bibinfo{person}{Lukasz Kaiser}, {and} \bibinfo{person}{Illia Polosukhin}.}
  \bibinfo{year}{2017}\natexlab{}.
\newblock \showarticletitle{Attention is all you need}.
\newblock \bibinfo{journal}{\emph{arXiv preprint arXiv:1706.03762}}
  (\bibinfo{year}{2017}).
\newblock


\bibitem[\protect\citeauthoryear{Wen, Aghajani, Nagy, Lanza, and Bavota}{Wen
  et~al\mbox{.}}{2021}]%
        {wen2021siri}
\bibfield{author}{\bibinfo{person}{Fengcai Wen}, \bibinfo{person}{Emad
  Aghajani}, \bibinfo{person}{Csaba Nagy}, \bibinfo{person}{Michele Lanza},
  {and} \bibinfo{person}{Gabriele Bavota}.} \bibinfo{year}{2021}\natexlab{}.
\newblock \showarticletitle{Siri, Write the Next Method}. In
  \bibinfo{booktitle}{\emph{2021 IEEE/ACM 43rd International Conference on
  Software Engineering (ICSE)}}. IEEE, \bibinfo{pages}{138--149}.
\newblock


\bibitem[\protect\citeauthoryear{Yang, Jiang, Gu, Sun, Gao, and Liu}{Yang
  et~al\mbox{.}}{2017}]%
        {yang2017language}
\bibfield{author}{\bibinfo{person}{Yixiao Yang}, \bibinfo{person}{Yu Jiang},
  \bibinfo{person}{Ming Gu}, \bibinfo{person}{Jiaguang Sun},
  \bibinfo{person}{Jian Gao}, {and} \bibinfo{person}{Han Liu}.}
  \bibinfo{year}{2017}\natexlab{}.
\newblock \showarticletitle{A language model for statements of software code}.
  In \bibinfo{booktitle}{\emph{2017 32nd IEEE/ACM International Conference on
  Automated Software Engineering (ASE)}}. IEEE, \bibinfo{pages}{682--687}.
\newblock


\bibitem[\protect\citeauthoryear{Zhang and Yang}{Zhang and Yang}{2021}]%
        {zhang2021survey}
\bibfield{author}{\bibinfo{person}{Yu Zhang} {and} \bibinfo{person}{Qiang
  Yang}.} \bibinfo{year}{2021}\natexlab{}.
\newblock \showarticletitle{A survey on multi-task learning}.
\newblock \bibinfo{journal}{\emph{IEEE Transactions on Knowledge and Data
  Engineering}} (\bibinfo{year}{2021}).
\newblock


\end{thebibliography}

\end{document}